\documentclass[aps,epsfig]{revtex4}
\usepackage{graphicx}

\begin{document}
\def\be{\begin{equation}}
\def\ee{\end{equation}}
\def\bfi{\begin{figure}}
\def\efi{\end{figure}}
\def\bea{\begin{eqnarray}}
\def\eea{\end{eqnarray}}
\newcommand{\ket}[1]{\vert#1\rangle}
\newcommand{\bra}[1]{\langle#1\vert}
\newcommand{\braket}[2]{\langle #1 \vert #2 \rangle}
\newcommand{\ketbra}[2]{\vert #1 \rangle  \langle #2 \vert}
\def\id{I}
\title{Nonlinear response and fluctuation dissipation relations}

\author{Eugenio Lippiello$^\dag$}
\affiliation{Dipartimento di Scienze Fisiche, Universit\'a di Napoli
``Federico II'', 80125 Napoli, Italy.}
\author{Federico Corberi$^\ddag$, Alessandro Sarracino$^\P$
and Marco Zannetti$^\S$}
\affiliation{Dipartimento di Matematica ed Informatica
via Ponte don Melillo, Universit\`a di Salerno, 84084 Fisciano (SA), Italy}

\begin{abstract}
A unified derivation of
the off equilibrium fluctuation dissipation relations
(FDR) is given for Ising and continous spins to arbitrary order,
within the framework of Markovian stochastic dynamics.
Knowledge of the FDR allows to develop zero field
algorithms for the efficient numerical computation of the response functions.
Two applications are presented. In the first one, the problem of probing for
the existence of a growing cooperative length scale is considered in those cases, like
in glassy systems, where the linear FDR is of no use. 
The effectiveness of an appropriate second order FDR is illustrated in the test case of the
Edwards-Anderson spin glass in one and two dimensions. In the second one, the important problem of the
definition of an off equilibrium effective temperature through the nonlinear
FDR is considered. It is shown that, in the case of coarsening systems, the effective temperature
derived from the second order FDR is consistent with the one obtained from the linear FDR.

\end{abstract}

\maketitle

\dag lippiello@sa.infn.it
\ddag corberi@sa.infn.it
\P sarracino@sa.infn.it
\S zannetti@sa.infn.it

\vspace{1cm}


\section{Introduction}

The statistical mechanics of systems out of equilibrium is a rapidly evolving subject, due to
the intensive research in the slow relaxation phenomena arising in
several different contexts, such as coarsening systems, glassy and granular materials,
colloidal systems etc. Understanding the basic mechanism underlying the slow relaxation
is an issue of major importance. In particular, a key question is whether the large time
scales are due to cooperative effects on large length scales. For non disordered
coarsening systems this is certainly the case, since the observed power law relaxation can be
directly related to the growth of the time dependent correlation length, or domain size~\cite{Bray}.
In the case of disordered or glassy systems the establishment of such a connection is
much more problematic, due to the difficulty of pinpointing the observables fit to the
task.

The use of the nonlinear susceptibilities has been advocated ~\cite{Huse,biroli,science} 
as experimental
or numerical probe apt to detect the heterogeneous character of the glassy relaxation and,
possibly, to uncover the existence of the growing length scale responsible of
the slowing down of the relaxation.
However, this requires to establish clearly the relationship between
non linear susceptibilities and correlation functions, in order to make sure what
actually do the nonlinear susceptibilities probe. In other words, the
problem of the derivation of the nonlinear fluctuation dissipation relations (FDR)
in the out of equilibrium regime needs to be addressed. As a matter of fact, this has been
one of the most fruitful lines of investigation in the field of slow relaxation,
although mostly limited to the domain of linear response~\cite{Cugl-review,CKP}.

In this paper we approach the problem on fairly general grounds.
Within the framework of Markovian stochastic evolution, we bring to the fore
the structural elements which are common to discrete and continous spins. We develop
the formal apparatus necessary for a unified derivation of the FDR in the two cases
and to arbitrary order. We also show, 
expanding on the work of Semerjian et al.~\cite{semerjian}, 
how the nonlinear FDR of arbitrary order can be derived from a fluctuation principle~\cite{Ritort} also
in the off equilibrium regime. This allows
to regard the FDR as a manifestation of the constraint imposed on the dynamics by
the requirement of microscopic reversibility.

The immediate application of the FDR is in the development of algorithms for the
computation of the response functions without the imposistion of an external perturbation,
the so called zero field algorithms. The numerical advantages of a zero field
algorithm are remarkable. These have been illustrated and discussed in detail,
in the linear case, in a recent
paper~\cite{lippiello05}. In the present paper we apply the zero field algorithm
to the computation of the nonlinear response functions. We consider two cases,
where knowledge of a nonlinear response function is required.
The first one arises in the search for a growing length scale in the context of glassy
systems, as mentioned above. The presence of quenched or self-induced disorder
makes the linear response function short ranged, compelling
to resort to the nonlinear ones.
The second one deals with the extension of the effective temperature concept~\cite{Peliti} to
nonlinear order. This is an important and difficult problem. Here, we consider it
in the context of non disordered coarsening systems, showing the consistency
of the effective temperatures derived from the linear and the second order FDR.

The paper is organised as follows: the formal developments concerning the time evolution,
the response functions, the FDR and the fluctuation principle are presented in
sections 2, 3, 4 and 5, respectively. In section 6  the problem of the detection of a 
cooperative length trough a nonlinear
susceptibility is addressed, while the effective temperature to nonlinear order
is treated in section 7. Concluding remarks are presented in section 8.

\section{Formalism and time evolution}
\label{2}

Let us consider a system in contact with a thermal reservoir, whose microscopic
states are the configurations $\sigma = [\sigma_i]$ of the $N$ degrees
of freedom, discrete or continous, placed on the discrete set of sites $i=1,..,N$. 
Assuming a Markovian stochastic dynamics,
the time evolution of the system is fully specified once the probability 
distribution $P(\sigma,t_0)$ at some initial time $t_0$ is given, together
with the transition probability $P(\sigma, t|\sigma^{\prime}, t^{\prime})$
for any pair of times $t_0 \leq t^{\prime} \leq t$.
Observables, or random variables, are functions $A(\sigma)$ defined 
over the phase space of the microscopic states, whose expectations
are given by
\be
\langle A(t) \rangle = \sum_{\sigma}A(\sigma)P(\sigma,t)
\label{form.1}
\ee
where $P(\sigma,t) = \sum_{\sigma^{\prime}}P(\sigma, t|\sigma^{\prime}, t_0)P(\sigma^{\prime},t_0)$.

In order to keep the notation compact, it is convenient to switch to
the operator formalism~\cite{KK}, whereby
the microscopic states introduced above are the set of labels of the basis vectors
\be
\ket{\sigma} = \bigotimes \ket{\sigma_i}, \;\;\;i=1,...N
\label{TE.1}
\ee
of a vector space. The single site states obey the orthonormality relation
$ \braket{\sigma_i}{\sigma^{\prime}_j}= \delta_{\sigma_i,\sigma^{\prime}_j}\delta_{i,j}$.
With this notation, the probability distributions $P(\sigma,t)$ become the time dependent vectors $\ket{P(t)}$
such that
\be
P(\sigma,t) = \braket{\sigma}{P(t)}
\label{TE.2}
\ee
and the transition probability is associated to the propagator of the process $\hat{P}(t|t^{\prime})$,
which is the operator whose matrix elements are given by
\be
P(\sigma,t|\sigma^{\prime},t^{\prime}) = \bra{\sigma}\hat{P}(t|t^{\prime})\ket{\sigma^{\prime}}.
\label{form.2}
\ee
For stochastically continous processes, through the first order expansion
\be
\hat{P}(t+\Delta t|t)= \hat{\id} + \hat{W}(t)\Delta t + {\cal O}(\Delta t^2)
\label{form.3}
\ee 
where $\hat{\id}$ is the identity operator, there remains defined the generator of the process $\hat{W}(t)$. 
Then, the pair $\ket{P(t_0)},\hat{W}(t)$ contains all the information on the process,
since the propagator can be written as
\be
\hat{P}(t| t^{\prime}) =  {\cal T}\left ( e^{\int_{t^{\prime}}^t ds \hat{W}(s)} \right)
\label{TE.4}
\ee
where ${\cal T}$ is the time ordering operator. In differential form this is equivalent to
the equations
\be
{\partial \over \partial t}\hat{P}(t| t^{\prime}) = \hat{W}(t)\hat{P}(t| t^{\prime})
\label{TE.5}
\ee
\be
{\partial \over \partial t^{\prime}}\hat{P}(t| t^{\prime}) = -\hat{P}(t| t^{\prime})\hat{W}(t^{\prime})
\label {TE.6}
\ee
from the first one of which follows the equation of motion of the state vector
\be
{\partial \over \partial t}\ket{P(t)}= \hat{W}(t)\ket{P(t)}.
\label{TE.8bis}
\ee
Notice that the normalization of probabilities imposes on the generator the condition
\be
\bra{-}\hat{W}(t)=0
\label {TE.7}
\ee
where $\bra{-} = \sum_{\sigma}\bra{\sigma}$ is called the flat vector.
We shall assume that at each instant of time the generator satisfies
the detailed balance condition
\be
e^{\beta \hat{{\cal H}}(t)}\hat{W}(t)e^{-\beta \hat{{\cal H}}(t)} = \hat{W}^{\dagger}(t)
\label{TE.9}
\ee
where $\hat{{\cal H}}(t)$ denotes the Hamiltonian of the system, with a possible time dependence.
This implies that the istantaneous Gibbs state
\be
\ket{P(t)}_{\beta}= {1 \over Z(t)}e^{-\beta \hat{{\cal H}}(t)}\ket{-}
\label{TE.10}
\ee
is an invariant state in the sense that
\be
\hat{W}(t)\ket{P(t)}_{\beta} = 0.
\label {TE.11}
\ee
The time dependent partition function is given by $Z(t) = \bra{-}e^{-\beta \hat{{\cal H}}(t)}\ket{-}$,
as it follows from the normalization of the state $\braket{-}{P(t)}_{\beta} = 1$. From now on we shall adopt the
notation $\ket{ \cdot }_{\beta}$ for the Gibbs states.

Notice that Eq.~(\ref{form.1}) requires that observables correspond to diagonal operators.
Each random function $A(\sigma)$ is mapped into the operator $\hat{A}=A(\hat{\sigma})$, where
the operators $\hat{\sigma}_{i}$ are
defined by 
\be
\hat{\sigma}_{i}\ket{\sigma}=\sigma_{i}\ket{\sigma}
\label{form.10}
\ee
in such a way that the expectation~(\ref{form.1}) can be written as
\be
\langle \hat{A}(t) \rangle = \bra{-}\hat{A}\ket{P(t)}.
\label{form.4}
\ee
If $\hat{W}$ and $\ket{P}_{\beta}$ are time independent, one can show that
the multi-time expectations of observables in the stationary state obey the Onsager relation~\cite{nota1}. 
Namely, if $\hat{A}_1,\hat{A}_2,...,\hat{A}_n$ is a set of observables and
$t_n \geq t_{n-1}...\geq t_1$ an ordered sequence of instants of time, then
\begin{eqnarray}
& & \langle \hat{A}_n(t_n)\hat{A}_{n-1}(t_{n-1})...\hat{A}_2(t_2)\hat{A}_1(t_1)\rangle_{\beta} = \nonumber \\
& & \langle \hat{A}_1(t_n) \hat{A}_2(t_n-(t_{2}-t_1))...\hat{A}_{n-1}(t_1+(t_{n}-t_{n-1}))\hat{A}_n(t_1)\rangle_{\beta}
\label{OR12}
\end{eqnarray}
where
\begin{eqnarray}
& & \langle \hat{A}_n(t_n)\hat{A}_{n-1}(t_{n-1})...\hat{A}_2(t_2)\hat{A}_1(t_1)\rangle_{\beta} = \nonumber \\
& & \bra{-} \hat{A}_n \hat{P}(t_n|t_{n-1})\hat{A}_{n-1}...
\hat{A}_2\hat{P}(t_2|t_1)\hat{A}_1 \ket{P}_{\beta}.
\label{exp.1}
\end{eqnarray}
Finally, using Eqs.~(\ref{TE.5}) and~(\ref{TE.6}), 
it is straightforward to show that the time derivative of a multi-time expectation is given by
\begin{eqnarray}
& & {\partial \over \partial t_k}\langle \hat{A}_n(t_n)..\hat{A}_k(t_k)..\hat{A}_1(t_1)\rangle = \nonumber \\
& & \langle \hat{A}_n(t_n)..[\hat{A}_k,\hat{W}(t_k)](t_k)..\hat{A}_1(t_1)\rangle
\label{comm.01}
\end{eqnarray}
namely, the time derivative in front of the expectation amounts to the insertion, at the time $t_k$, of the
commutator $[\hat{A}_k,\hat{W}(t_k)]$ inside the expectation. This applies for all $k=1,..,n$.
In particular, in the case of a single observable
\be
{\partial \over \partial t}\langle \hat{A}(t)\rangle = \bra{-}\hat{A}\hat{W}(t)\ket{P(t)}
= \bra{-}[\hat{A},\hat{W}(t)]\ket{P(t)}
\label{comm.02}
\ee
where the second equality is a consequence of Eq.~(\ref{TE.7}).
It should be noticed that, in general, $[\hat{A},\hat{W}(t)]$ is not an observable.

\section{Response functions}

Let us assume that the generator $\hat{W}(t)$ depends on time through an external field $h_i(t)$.
Then, upon varying $h_i(t)$, there remains defined a family of stochastic processes and one
may ask whether these stochastic processes are related one to the other. In particular,
one would like to know whether the process with the generic $h_i(t)$ can be reconstructed
from the properties of the {\it unperturbed} process, the one with the particular
choice $h_i(t) \equiv 0$. In order to answer the question, let us start from the
statement that all the information in the process, with a given $h_i(t)$,
is contained in the full hierarchy of the time dependent moments 
\be
M_{i_1,...,i_n}(t_1,...,t_n,[h_i(t^{\prime})]) = 
\langle \hat{\sigma}_{i_1}(t_1)...\hat{\sigma}_{i_n}(t_n)\rangle
\label{respf.1}
\ee
each of which is a functional of $h_i(t)$. Assuming analiticity, one can 
write the formal expansion
\begin{eqnarray}
& & M_{i_1..i_n}(t_1,..,t_n,[h_i(t^{\prime})]) =  M_{i_1..i_n}(t_1,..,t_n) + \\ \nonumber
&  & \sum_{m=1}^{\infty} {1 \over m!}\sum_{j_1..j_m}
\int_{t_w}^t dt^{\prime}_1 ...\int_{t_w}^t dt^{\prime}_m \; 
R^{(n,m)}_{i_1..i_n;j_1..j_m}(t_1,..,t_n;t^{\prime}_1,..,t^{\prime}_m)h_{j_1}(t^{\prime}_1)...h_{j_m}(t^{\prime}_m)
\label{respf.2}
\end{eqnarray}
where  $M_{i_1..i_n}(t_1,..,t_n)$ is the unperturbed moment, $(t_w,t)$ is the time interval
over which the action of the external field is considered, and
\be
R^{(n,m)}_{i_1..i_n;j_1..j_m}(t_1,..,t_n;t^{\prime}_1,..,t^{\prime}_m)
= \left . {\delta^m  M_{i_1..i_n}(t_1,..,t_n,[h_i(t^{\prime})]) \over
\delta h_{j_1}(t^{\prime}_1)... \delta h_{j_m}(t^{\prime}_m)} \right |_{h=0} 
\label{respf.3}
\ee
is the $m$-th order response function of the $n$-th moment. Therefore, the question asked above can
be positively answered if the response functions can be expressed
in terms of quantities computable in the unperturbed process, namely if 
the FDR of arbitrary order can be obtained. 

Without loss of generality, we limit ourselves to work out the response functions for the first moment.
For the responses of the higher moments there is nothing conceptually different,
just the formalism gets more involved. As a matter of fact, in 
section~\ref{length} we shall deal with the second order response
of the second moment. Coming back to the first moment, from
\be
M_{i}(t,[h_j(t^{\prime})]) = \bra{-} \hat{\sigma}_i \hat{P}_h(t| t_w)\ket{P(t_w)}
\label{respf.4}
\ee
where $\hat{P}_h(t| t_w)$ is the propagator in the presence of the field, follows
\be
R^{(1,m)}_{i;j_1..j_m}(t,t_1,..,t_m) =  
\bra{-} \hat{\sigma}_i \left . {\delta^m \hat{P}_h(t| t_w) \over
\delta h_{j_1}(t_1)... \delta h_{j_m}(t_m)} \right |_{h=0} \ket{P(t_w)}
\label{respf.5} 
\ee
which obviously vanishes if $t_j \not \in (t_w,t)$ for any $j=1,..,m$.
Let us write explicitely the first two derivatives 
\begin{eqnarray}
{\delta \hat{P}_h(t| t_w) \over \delta h_{j_1}(t_1)} & = & 
{\cal T}\left [ e^{\int_{t_w}^t ds \hat{W}(s)}
{\partial \hat{W}(t_1) \over \partial h_{j_1}(t_1)} \right]\nonumber \\
 & = & \hat{P}_h(t| t_1){\partial \hat{W}(t_1) \over \partial h_{j_1}(t_1)} \hat{P}_h(t_1| t_w)
\label{respf.6}
\end{eqnarray}
and
\begin{eqnarray}
{\delta^2 \hat{P}_h(t| t_w) \over \delta h_{j_1}(t_1) \delta h_{j_2}(t_2)} & = &
{\cal T}\left \{ e^{\int_{t_w}^t ds \hat{W}(s)} \left [
{\partial \hat{W}(t_1) \over \partial h_{j_1}(t_1)} {\partial \hat{W}(t_2) \over \partial h_{j_2}(t_2)}
+{\partial^2 \hat{W}(t_1) \over \partial h_{j_1}^2(t_1)} \delta(12) \right ]
\right \} \nonumber \\
& = & \hat{P}_h(t| t_M){\partial \hat{W}(t_M) \over \partial h_{j_M}(t_M)} \hat{P}_h(t_M| t_m)
{\partial \hat{W}(t_m) \over \partial h_{j_m}(t_m)}\hat{P}_h(t_m| t_w) \nonumber \\
& + & \hat{P}_h(t| t_1){\partial^2 \hat{W}(t_1) \over \partial h_{j_1}^2(t_1)} \hat{P}_h(t_1| t_w)
\delta(12)
\label{respf.7}
\end{eqnarray}
where $t_M = \max(t_j)$, $t_m = \min(t_j)$, 
$j_M,j_m$ are the sites where the field acts at the times $t_M$ or $t_m$, respectively,
and $\delta(np) = \delta_{j_n,j_p}\delta(t_n-t_p)$. The third order derivative is written down 
in Appendix I.

In order to go further, we must specify how $\hat{W}(t)$ depends on the field $h_{i}(t)$ and that is
where the distinction between discrete spins and continous spins comes in.
In the following, in order to keep the derivation as simple as possible we shall specialize to
the case of Ising spins with single spin-flip dynamics. The generalization to $q$-state
spins (e.g. clock models) and to dinamical rules with  conservation laws, such as Kawasaki spin-exchange,
turns out to be straightforward.

\subsection{Ising spins}
\label{IS}

For Ising spins the state vectors $\ket{\sigma_i=\pm 1}$
are represented by the column vectors
\begin{displaymath}
\ket{\sigma_i=1}=\left(
\begin{array}{c}
  1 \\
  0 \\
  \end{array}
\right), \;\;\;\;
\ket{\sigma_i=-1}=\left(
\begin{array}{c}
  0 \\
  1 \\
  \end{array}
\right).
\end{displaymath}
A generator of single spin flip dynamics is of the type
\be
\hat{W} = {1 \over N}\sum_{i=1}^N \hat{W}_i
\label{TE.12}
\ee
where $\hat{W}_i$ has non vanishing matrix elements only between states which differ for the value
of $\sigma_i$. The form of $\hat{W}_i$ is obtained imposing
the detailed balance condition~(\ref{TE.9}). Singling out the site $i$,
the Hamiltonian can be written in the form
\be
\hat{{\cal H}}(t)= {\cal H}([\hat{\sigma}^z]_i,t) + \{h^W([\hat{\sigma}^z]_i) - h_i(t)\}\hat{\sigma}^z_i,
\label{TE.13}
\ee
where $\hat{\sigma}^z_i$ is the $z$ Pauli matrix
\begin{displaymath}
\hat{\sigma}^z_i=\left(
\begin{array}{cc}
  1 & 0 \\
  0 & -1\\
  \end{array}
\right),
\end{displaymath}
$[\hat{\sigma}^z]_i$ stands for the set of all spins except for the one on
the $i$-th site and $\hat{h}^W_i=h^W([\hat{\sigma}^z]_i)$ is
the Weiss field on the site $i$.
Inserting into the detailed balance condition~(\ref{TE.9}), one finds the generator of the 
Glauber type~\cite{Glauber} 
\be
\hat{W}_i(t) = (\hat{\sigma}^x_i - \hat{\id})
e^{\beta[\hat{h}^W_i-h_i(t)]\hat{\sigma}^z_i}
\label{TE.14}
\ee
where $\hat{\sigma}^x_i$ is the $x$ Pauli matrix
\begin{displaymath}
\hat{\sigma}^x_i=\left(
\begin{array}{cc}
  0 & 1 \\
  1 & 0\\
  \end{array}
\right).
\end{displaymath}
Hence, the derivatives are given by
\be
{\partial^n \hat{W}(t_1) \over \partial h_{j_1}^n(t_1)} =
(-\beta)^n \hat{W}_{j_1}(t_1)(\hat{\sigma}^z_{j_1})^n
\label{RF.05}
\ee
with
\be
(\hat{\sigma}^z_{j_1})^n  = \left \{ \begin{array}{ll}
        \hat{\id}, \;\;$for$ \;\; n \;\; $even$ \\
        \hat{\sigma}^z_{j_1},  \;\; $for$ \;\; n \;\; $odd$.
        \end{array}
        \right .
        \label{RF.06}
        \ee
For later reference, let us write here the identity
\be
\hat{W}_i \hat{\sigma}^z_i  = {1 \over 2}[\hat{W}_i, \hat{\sigma}^z_i] + {1 \over 2} \{\hat{W}_i , \hat{\sigma}^z_i \}
\label{IS.01}
\ee
obtained by adding the commutator and the anticommutator. It is straightforward to verify
that the anticommutator is a diagonal operator and, therefore, is an observable which we will
denote by
\be
\hat{B}_i(t) = \{\hat{\sigma}^z_i,\hat{W}(t)\}. 
\label{B.1}
\ee
Due to Eq.~(\ref{TE.7}), the average of the left hand side of Eq.~(\ref{IS.01}) vanishes and
Eq.~(\ref{comm.02}) for $\hat{\sigma}^z_i$ can be rewritten as
\be
{\partial \over \partial t}\langle \hat{\sigma}^z_i(t)\rangle = \langle \hat{B}_i(t) \rangle.
\label{comm.02bis}
\ee

\subsection{Continuous spins}

In order to avoid confusion, here we shall denote by $\varphi=[\varphi_{i}]$ the set of the $N$ continuous
variables. Let us assume an equation of motion of the Langevin type 
\be {\partial \over \partial t}\varphi_i(t)= B_i(t) +
\eta_i(t) \label{CS1} 
\ee 
where the drift $B_i$ is related to the Hamiltonian, or free energy functional ${\cal H}[\varphi]$, by
\be B_i(t) =
- {\partial {\cal H}[\varphi(t)] \over \partial \varphi_i}
\label{CS2} 
\ee 
$\eta_i(t)$ is a gaussian white noise with expectations 
\be
\langle  \eta_i(t) \rangle = 0, \;\;\;\;\; \langle  \eta_i(t)
\eta_j(t^{\prime})\rangle = 2T \delta_{i,j}\delta(t-t^{\prime})
\label{CS3} 
\ee 
and $T$ is the temperature of the thermal reservoir. 
The generator of the corresponding Markov process is the Fokker-Planck operator
\be
\hat{W}^{FP}=\sum_i\hat{W}_i^{FP}
\label{CS4}
\ee
where
\be
\hat{W}_i^{FP} = -\{ T\hat{p}_i^2 +iB_i[\hat{\varphi}]\hat{p}_i + D_i[\hat{\varphi}] \},
\label{CS5}
\ee
$B_i[\hat{\varphi}]$ is defined through Eq.~(\ref{CS2}) and
\be
D_i[\hat{\varphi}] = - \left . {\partial^2 {\cal H}[\varphi] \over \partial \varphi_i^2}
\right |_{\varphi = \hat{\varphi}}.
\label{CS6}
\ee
The conjugated operators $\hat{\varphi}_i$ and  $\hat{p}_i$, defined by
\be
\bra{\varphi^{\prime}}\hat{\varphi}_i \ket{\varphi} = \varphi_i \braket{\varphi^{\prime}}{\varphi}
\label{CS8}
\ee
\be
\bra{\varphi^{\prime}}\hat{p}_i \ket{\varphi} = -i {\partial \over \partial \varphi_i}
\braket{\varphi^{\prime}}{\varphi}
\label{CS9}
\ee
obey the commutation relation
\be
[\hat{\varphi}_i,\hat{p}_j]=i \delta_{i,j}
\label{CS10}
\ee
and satisfy the equalities
\be
\bra{-} \hat{p}_i = 0
\label{norm.1}
\ee
\be
\bra{-} \hat{W}_i^{FP} = 0.
\label{norm.2}
\ee
The external field enters the free energy functional linearly
\be
{\cal H}_h[\varphi,t]= {\cal H}[\varphi] - \sum_i h_i(t)\varphi_i
\label{CS11}
\ee
yielding
\be
B_{h,i}[\varphi,t] = B_{i}[\varphi] +  h_i(t),
\label{CS12}
\ee
while $D_i[\varphi]$ remains unaltered. Therefore, the generator is changed into
\be
\hat{W}_{h,i}^{FP}=\hat{W}_i^{FP} -i h_i(t)\hat{p}_i
\label{CS13}
\ee
and the derivatives are given by
\be
{\partial^n \hat{W}_{h,i}^{FP} \over \partial  h^n_i(t)}  = \left \{ \begin{array}{ll}
         -i \hat{p}_i, \;\;$for$ \;\; n=1 \\
        0,  \;\; $for$ \;\; n>1.
        \end{array}
        \right .
        \label{CS14}
        \ee
Obviously, the identity~(\ref{IS.01}) holds also in this case 
\be
\hat{W}_i^{FP} \hat{\varphi}_i  = 
{1 \over 2}[\hat{W}_i^{FP}, \hat{\varphi}_i] + {1 \over 2} \{\hat{W}_i^{FP} , \hat{\varphi}_i \}
\label{IS.01bis}
\ee
and, using the definiton~(\ref{CS5}) together with
the commutation relation~(\ref{CS10}), it is not difficul to rewrite it in the more convenient form
\be
i \hat{p}_i = {\beta \over 2} \left \{[\hat{W}_i^{FP},\hat{\varphi}_i]
+  B_i[\hat{\varphi}] \right \}.
\label{CS10.4}
\ee
Again, since the average of the left hand side vanishes, we get the analogous of Eq.~(\ref{comm.02bis})
\be
{\partial \over \partial t}\langle \hat{\varphi}_i(t)\rangle = \langle  B_i[\hat{\varphi}](t) \rangle
\label{comm.02ter}
\ee
showing that the anticommutator~(\ref{B.1}) for Ising spins and the drift~(\ref{CS2}) for continous spins 
play the same role in the evolution, thus justifying the
choice of the same notation.

\section{Fluctuation dissipation relations}

Let us now return to the general treatment, valid for discrete and continous spins
alike, with the operator $\hat{\sigma}$ standing either for $\hat{\sigma}^z$ or
for $\hat{\varphi}$, depending on the context. Which is the case, will be specified
whenever necessary.

Comparing the left hand sides of Eqs.~(\ref{IS.01})and~(\ref{CS10.4}) with
the first derivatives~(\ref{RF.05}) and~(\ref{CS14}), we can write the basic equation in this paper
\be
{\partial \hat{W} \over \partial h_{i}} =
{-\beta \over 2} \left ( [\hat{W}_{i},\hat{\sigma}_{i})] + B_i[\hat{\sigma}] \right )
\label{nw.2}
\ee
where, in the case of continous spins, the superscript FP on the Fokker-Planck operator has been dropped
and, as specified above, $\hat{B}_i$ stands for
\be
\hat{B}_i  = \left \{ \begin{array}{ll}
        \{ \hat{\sigma}_{i},\hat{W}_i \}, \;\;$for Ising spins$  \\
        B_i[\hat{\varphi}],  \;\; $for continous spins$.
        \end{array}
        \right .
        \label{nw.3ter}
        \ee
We are now in the position to derive the FDR, by going through the
following steps:

\begin{enumerate}

\item according to Eq.~(\ref{respf.5}), the response function is obtained by
appropriate insertions of derivatives of the generator in between propagators,
as exemplified in Eqs.~(\ref{respf.6}) and~(\ref{respf.7}) 

\item according to Eq.~(\ref{nw.2}), each first derivative of $\hat{W}$ can be repalced by the sum
of the commutar and the drift operator $\hat{B}_i$

\item according to Eq.~(\ref{comm.01}), the insertion of the commutator amounts 
to a time derivative acting in front of the average.

\end{enumerate}

The above steps exaust all there is to do in the continous spin case, since there are
no derivatives of the generator of order higher than the first. Higher derivatives do,
instead, appear in the Ising case producing singular terms. In order to see how the procedure works
in practice, let us carry out the computation up to second order. The explicit result
for the third order response function is presented in Appendix I.

\subsection{Linear FDR}

Let us begin with the simplest case of the linear response function.
From Eqs.~(\ref{respf.5}) and~(\ref{respf.6}) follows
\be
R^{(1,1)}_{i;j_1}(t,t_1) =
\bra{-} \hat{\sigma}_i \hat{P}(t| t_1)  \left . {\partial \hat{W}(t_1) \over \partial h_{j_1}(t_1)} \right |_{h=0}
\hat{P}(t_1| t_w)\ket{P(t_w)}
\label{lrsp1}
\ee
and, using Eqs.~(\ref{nw.2}) and~(\ref{comm.01}), this becomes the linear FDR 
\be
R^{(1,1)}_{i;j_1}(t,t_1) =  {\beta \over 2} \left [{\partial \over \partial t_1} M_{ij_1}(t,t_1)
-  \langle \hat{\sigma}_i (t) \hat{B}_{j_1}(t_1) \rangle \right ]
\label{RF14}
\ee
due to the appearence of unperturbed correlation functions of observables in the right hand side.
This result has been known for some time for continous spins,
see e.g. Ref.~\cite{CKP}, while for Ising spins has been derived for the first time
in Ref.~\cite{lippiello05}, where it has been exploited to develop the zero field
algorithm for the computation of $R^{(1,1)}_{i;j_1}(t,t_1)$, mentioned in the Introduction.
 
If the system is in equilibrium, the averages in the right hand side are equilibrium averages and,
invoking the Onsager relation~(\ref{OR12}), one gets
\be
\langle \hat{\sigma}^z_i (t) \hat{B}_{j_1}(t_1) \rangle_{\beta} =
\bra{-} [\hat{\sigma}_{j_1},  \hat{W}]\hat{P}(t| t_1)\hat{\sigma}_{i}\ket{P}_{\beta}=
-{\partial \over \partial t_1} M_{ij_1}(t,t_1)
\label{RF17}
\ee
after using space and time translation invariance in the last equality,
Eqs.~(\ref{IS.01}) or~(\ref{CS10.4}) to eliminate $\hat{B}_{j_1}$, together with
the normalization conditions~(\ref{TE.7}) or~(\ref{norm.1}).
Inserting this into Eq.~(\ref{RF14}), the equilibrium fluctuation dissipation theorem (FDT) is recovered
\be
R^{(1,1)}_{ij_1}(t,t_1) =  \beta {\partial \over \partial t_1}C_{ij_1}(t,t_1)
\label{RF18}
\ee
where we have introduced the pair correlation function
\be
C_{ij_1}(t,t_1) =  M_{ij_1}(t,t_1) -  M_{i}(t) M_{j_1}(t_1)
\label{corr.1}
\ee
since it is clear that in equilibrium the one time quantities 
are time independent and the time derivative of $C_{ij_1}(t,t_1)$ coincides with that
of $M_{ij_1}(t,t_1)$.

\subsection{Second order FDR}
\label{two_kicks}

From Eqs.~(\ref{respf.5}) and~(\ref{respf.7}), the second order, 
or {\it two kicks} nonlinear response function, is given by
\begin{eqnarray}
& & R^{(1,2)}_{i;j_1j_2}(t,t_1,t_2) = \nonumber \\
& & \bra{-} \hat{\sigma}_i \hat{P}(t| t_M) \left . {\partial \hat{W}(t_M) \over \partial h_{j_M}(t_M)} \right |_{h=0} 
\hat{P}(t_M| t_m) \left .
{\partial \hat{W}(t_m) \over \partial h_{j_m}(t_m)} \right |_{h=0}  \hat{P}_h(t_m| t_w) \ket{P(t_w)} \nonumber \\
& + & \bra{-}\hat{\sigma}_i 
\hat{P}(t| t_1) \left . {\partial^2 \hat{W}(t_1) \over \partial h_{j_1}^2(t_1)}\right |_{h=0}  
\hat{P}(t_1| t_w)\ket{P(t_w)} \delta(12).
\label{NL3}
\end{eqnarray}
Substituting, next, Eq.~(\ref{nw.2}) for the first derivatives and the identity
\be
\hat{W}_{i} = {1 \over 2}\hat{\sigma}^z_i \hat{B}_i 
+ {1 \over 2}\hat{\sigma}^z_i [\hat{\sigma}^z_i, \hat{W}_{i}] 
\label{ident}
\ee
for the manipulation of the singular term, eventually
one finds the second order FDR
\begin{eqnarray}
R^{(1,2)}_{i;j_1j_2}(t,t_1,t_2) & = &  ( \beta /2 )^2
\Big \{ {\partial \over \partial t_M}{\partial \over \partial t_m} 
M_{ij_Mj_m}(t,t_M,t_m)  \nonumber \\
& - & {\partial \over \partial t_M}
\langle \hat{\sigma}_i (t)\hat{\sigma}_{j_M}(t_M)
\hat{B}_{j_m}(t_m)\rangle  \nonumber \\
& - & {\partial \over \partial t_m}
\langle \hat{\sigma}_i (t)\hat{B}_{j_M}(t_M)
\hat{\sigma}_{j_m}(t_m)\rangle  \nonumber \\
& + &  \langle \hat{\sigma}_i (t)\hat{B}_{j_M}(t_M)
\hat{B}_{j_m}(t_m)\rangle   \Big \} \nonumber \\
& + & (\beta^2/2) \Big \{ \langle \hat{\sigma}^z_i (t) \hat{\sigma}^z_{j_m}(t_M)\hat{B}_{j_m}(t_m) \rangle
 \nonumber \\
& + &   {\partial \over \partial t_m} M_{ij_mj_m}(t,t_M,t_m) \Big \}
\delta_{j_M,j_m} \delta(t_M - t_m) \nonumber \\
\label{NL8bis}
\end{eqnarray}
where, it is worth recalling, the last singular contribution in the braces is present
only for Ising spins.

Again, at stationarity the Onsager relation can be used
to eliminate the $\hat{B}$'s entering with the shortest time in favor of time derivatives. In so doing,
the second and the fourth contribution in the right hand side become identical to the first
and the third one, respectively. Furthermore, in the stationary state the last contribution
in the right hand side vanishes, eventually yielding what we may call the second order FDT
\begin{eqnarray}
R^{(1,2)}_{i;j_1j_2}(t,t_1,t_2) & = &  ( \beta^2 /2 )
\left \{ {\partial \over \partial t_M}{\partial \over \partial t_m}
C_{ij_Mj_m}(t,t_M,t_m)  \right . \nonumber \\
& - & \left . {\partial \over \partial t_m}
\langle \hat{\sigma}_i (t)\hat{B}_{j_M}(t_M)
\hat{\sigma}_{j_m}(t_m)\rangle_{\beta} \right \}
\label{TK3}
\end{eqnarray}
where we have substituted the time derivative of the
moment with that of the correlation function.
Notice that there remains an ineliminable presence of $\hat{B}$ in the second
term in the right hand side, which carries the information on the specific
rule governing the time evolution. This is a distinctive feature of all FDR
of order higher than linear, making them less universal than the linear one,
which is $\hat{B}$ dependent only off equilibrium.

\section{Fluctuation principle}

In the following we show that the FDR obtained in the previous section arise
as a consequence of a fluctuation principle~\cite{Ritort}.
For convenience, the derivation is presented for the continous spin case, but it can be worked
out along the same lines also for Ising spins. The idea of a derivation
from the fluctuation principle was implemented by Semerjian et al.~\cite{semerjian} in the stationary case.
Here we extend the derivation to the more general off equilibrium case.

Let us call experimental
protocol the assigned time dependence of the external field $h(t)$ 
in some time interval $(t_0,t_F)$. Then, from the detailed balance condition
follows that the probability $P_{\beta}([\varphi(t)]|\varphi_0,[h(t)])$  of a path $[\varphi(t)]$, taking
place under the protocol  $[h(t)]$ and conditioned to the initial value $\varphi(t_0)=\varphi_0$ is related to
the probability of the reverse path
\be
\widetilde{\varphi}(t) = \varphi(\widetilde{t}),\;\;\; \widetilde{t} = t_F -t+ t_0
\label{FP.1}
\ee
under the reverse protocol $[\widetilde{h}(t)]$ and conditioned to $\widetilde{\varphi}(t_0)=\varphi_F$
by
\begin{eqnarray}
& & P_{\beta}([\varphi(t)]|\varphi_0,[h(t)])
\exp \left \{-\beta \int_{t_0}^{t_F} dt \; h(t)\dot{\varphi}(t) \right \}
\nonumber \\
& = & P_{\beta}([\widetilde{\varphi}(t)]|\varphi_F,[\widetilde{h}(t)])
\exp \left \{\beta [ {\cal H}_0(\varphi_0)
-{\cal H}_0(\varphi_F)] \right \}
\label{Ub.14}
\end{eqnarray}
where the subscript $\beta$ is there to remind that the evolution takes place
while the system is in contact with a single thermal reservoir at the inverse
temperature  $\beta$.
Multiplying both
sides by $\varphi_{i,F} P_I(\varphi_0)$, where $P_I(\varphi_0)$ is an arbitrary
probability distribution, and summing over the set ${\cal C}(t_0,t_F)$ of all paths in the interval $(t_0,t_F)$
one finds
\begin{eqnarray}
& & \int_{{\cal C}(t_0,t_F)}  d[\varphi(t)]
\varphi_{i,F}  P_{\beta}([\varphi(t)]|\varphi_0,[h(t)])
\exp \left \{-\beta \int_{t_0}^{t_F} dt \; h(t)\dot{\varphi}(t) \right \} P_I(\varphi_0) \nonumber \\
& = &  Z_{\beta,0}  \int_{{\cal C}(t_0,t_F)} d[\varphi(t)] P_I(\varphi_0)  e^{\beta  {\cal H}_0(\varphi_0)}
P_{\beta}([\widetilde{\varphi}(t)]|\varphi_F,[\widetilde{h}(t)])
\varphi_{i,F} P_{\beta,0}(\varphi_F) \nonumber \\
\label{Ub.15}
\end{eqnarray}
where  $P_{\beta,0}$ and $ Z_{\beta,0}$ denote the equilibrium distribution
and the corresponding partition function, in absence of the external field.
Hence, the above result can be rewritten more compactly as
\begin{eqnarray}
& & \left \langle \varphi_i(t_F)  \exp \left \{-\beta \int_{t_0}^{t_F} dt \; h(t)\dot{\varphi}(t) \right \}
\right \rangle_{I \rightarrow \beta,[h(t)]} \nonumber \\
& = &  Z_{\beta,0}  \left \langle P_I(\varphi(t_F)) e^{\beta  {\cal H}_0(\varphi(t_F))}  \varphi_i(t_0)
\right \rangle_{\beta,0 \rightarrow \beta,[\widetilde{h}(t)]}
\label{Ub.16}
\end{eqnarray}
where $\langle \cdot \rangle_{I \rightarrow \beta,[h(t)]}$ stands for the average in the process
starting with the initial
condition $P_I$, thereafter in contact with the thermal resorvoir $\beta$ and evolving with the protocol $[h(t)]$,
while  $\langle \cdot \rangle_{\beta,0 \rightarrow \beta,[\widetilde{h}(t)]}$ stands for the process in contact
with the thermal resorvoir $\beta$, starting with the unperturbed equilibrium distribution
$P_{\beta,0}$ and evolving with the reverse protocol $[\widetilde{h}(t)]$.

The next step is to expand both sides in powers of $[h(t)]$  about $h(t) \equiv 0$
and to compare terms of the same order. This is done in Appendix II, obtaining
to zero order
\be
Z_{\beta,0} \left \langle P_I(\varphi(t_F)) e^{\beta  {\cal H}_0(\varphi(t_F))}  \varphi_i(t_0)
\right \rangle_{\beta,0} = \langle \varphi_i(t_F) \rangle_{I \rightarrow \beta,0}
\label{Ub.26}
\ee
where $\langle \cdot \rangle_{I \rightarrow \beta,0}$ stands for the average in the off-equilibrium process
starting with $P_I$ and evolving in contact with the thermal reservoir
in absence of the external field.
At higher orders one gets
\begin{eqnarray}
& & Z_{\beta,0}
\left . {\delta^n
\left \langle P_I(\varphi(t_F)) e^{\beta  {\cal H}_0(\varphi(t_F))}  \varphi_i(t_0)
\right \rangle_{\beta,0 \rightarrow \beta,[\widetilde{h}(t)]}  \over
\delta \widetilde{h}_{j_1}(\widetilde{t}_1)...\delta\widetilde{h}_{j_n}(\widetilde{t}_n)} 
\right |_{\widetilde{h}=0} = \nonumber \\
& & \sum_{p=0}^{n}(-\beta)^{n-p}
\sum_{{\cal P}(j_1,..,j_{n-p}|j_{n-p+1},..,j_n)}
\left . {\delta^p \langle \varphi_i(t_F)\dot{\varphi}_{j_1}(t_1)...\dot{\varphi}_{j_{n-p}}(t_{n-p})
\rangle_{I \rightarrow \beta,[h(t)]} \over \delta h_{j_{n-p+1}}(t_{n-p+1})...\delta h_{j_n}(t_n)}
\right |_{h=0} \nonumber \\
\label{Ub.27}
\end{eqnarray}
where the second summation is over all the distinct permutations

\noindent ${\cal P}(j_1,..,j_{n-p}|j_{n-p+1},..,j_n)$
between the two sets of indeces
$(j_1,...,j_{n-p})$ and $(j_{n-p+1},...,j_n)$.
For instance, at the first two orders the above formula reads
\begin{eqnarray}
& & Z_{\beta,0}
\left . {\delta
\left \langle P_I(\varphi(t_F)) e^{\beta  {\cal H}_0(\varphi(t_F))}  \varphi_i(t_0)
\right \rangle_{\beta,0\rightarrow \beta,[\widetilde{h}(t)]}  \over
\delta \widetilde{h}_{j_1}(\widetilde{t}_1)} \right |_{\widetilde{h}=0} = \nonumber \\
& - & \beta {\partial \over \partial t_1} \langle \varphi_i(t_F) \varphi_{j_1}(t_1)\rangle_{I \rightarrow \beta,0}
+ \left . {\delta \langle \varphi_i(t_F) \rangle_{I \rightarrow \beta,[h(t)]}
\over \delta h_{j_{1}}(t_{1})} \right |_{h=0}
\label{Ub.28}
\end{eqnarray}
and
\begin{eqnarray}
& & Z_{\beta,0}
\left . {\delta^2
\left \langle P_I(\varphi(t_F)) e^{\beta  {\cal H}_0(\varphi(t_F))}  \varphi_i(t_0)
\right \rangle_{\beta,0 \rightarrow \beta,[\widetilde{h}(t)]}  \over
\delta \widetilde{h}_{j_1}(\widetilde{t}_1)\delta\widetilde{h}_{j_2}(\widetilde{t}_2)} 
\right |_{\widetilde{h}=0} = \nonumber \\
& & \beta^2 {\partial \over \partial t_1} {\partial \over \partial t_2}
\langle \varphi_i(t_F) \varphi_{j_1}(t_1) \varphi_{j_2}(t_2) \rangle_{I \rightarrow \beta,0} \nonumber \\
& - & \beta {\partial \over \partial t_1}
\left . {\delta \langle \varphi_i(t_F) \varphi_{j_1}(t_1)\rangle_{I \rightarrow \beta,[h(t)]}
\over \delta h_{j_{2}}(t_{2})} \right |_{h=0}
- \beta {\partial \over \partial t_2}
\left . {\delta \langle \varphi_i(t_F) \varphi_{j_2}(t_2)\rangle_{I \rightarrow \beta,[h(t)]}
\over \delta h_{j_{1}}(t_{1})} \right |_{h=0} \nonumber \\
& + & \left . {\delta^2 \langle \varphi_i(t_F) \rangle_{I \rightarrow \beta,[h(t)]}
\over \delta h_{j_{1}}(t_{1}) \delta h_{j_2}(t_2)}
\right |_{h=0}.
\label{Ub.29}
\end{eqnarray}
From these two equations the off equilibrium  FDR~(\ref{RF14}) and~(\ref{NL8bis})
can be recovered. Here we show this explicitely in the linear case, the extension to
higher orders being straightforward.

Recalling the definition~(\ref{respf.3}) of the response functions and making
the replacements $t_F \rightarrow t$ and $t_0 \rightarrow t_w$, Eq.~(\ref{Ub.28})
can be rewritten as
\begin{eqnarray}
& & R^{(1,1)}_{i;j_1}(t,t_1) =
\beta {\partial \over \partial t_1} \langle \varphi_i(t) \varphi_{j_1}(t_1)\rangle_{I \rightarrow \beta,0} \nonumber \\
& + & Z_{\beta,0}
\left . {\delta
\left \langle P_I(\varphi(t)) e^{\beta  {\cal H}_0(\varphi(t))}  \varphi_i(t_w)
\right \rangle_{\beta,0 \rightarrow \beta,[\widetilde{h}(t)]}  \over
\delta \widetilde{h}_{j_1}(\widetilde{t}_1)} \right |_{\widetilde{h}=0}.
\label{Ub.30}
\end{eqnarray}
According to Eq.~(\ref{CS14}), the derivative with respect to $\widetilde{h}_{j_1}(\widetilde{t}_1)$
can be replaced by the insertion of
$-i\hat{p}_{j_1}(\widetilde{t}_1)$ and, using Eq.~(\ref{CS10.4}), the second term in the
right hand side can be rewritten as
\begin{eqnarray}
& & Z_{\beta,0}
\left . {\delta
\left \langle P_I(\varphi(t)) e^{\beta  {\cal H}_0(\varphi(t))}  \varphi_i(t_w)
\right \rangle_{\beta,0 \rightarrow \beta,[\widetilde{h}(t)]}  \over
\delta \widetilde{h}_{j_1}(\widetilde{t}_1)} \right |_{\widetilde{h}=0} = \nonumber \\
& &  Z_{\beta,0}{\beta \over 2} {\partial \over \partial \widetilde{t}_1}
\left \langle P_I(\varphi(t)) e^{\beta  {\cal H}_0(\varphi(t))}
\varphi_{j_1}(\widetilde{t}_1) \varphi_i(t_w) \right \rangle_{\beta,0} \nonumber \\
& - &  Z_{\beta,0}{\beta \over 2} \left \langle P_I(\varphi(t)) e^{\beta  {\cal H}_0(\varphi(t))}
B_{j_1}(\widetilde{t}_1) \varphi_i(t_w) \right \rangle_{\beta,0}.
\label{Ub.31}
\end{eqnarray}
Furthermore, since after setting to zero the external field the averages become
equilibrium averages, using the Onsager relation we get
\begin{eqnarray}
& & Z_{\beta,0}
\left . {\delta
\left \langle P_I(\varphi(t_F)) e^{\beta  {\cal H}_0(\varphi(t_F))}  \varphi_i(t_0)
\right \rangle_{\beta,0 \rightarrow \beta,[\widetilde{h}(t)]}  \over
\delta \widetilde{h}_{j_1}(\widetilde{t}_1)} \right |_{\widetilde{h}=0} = \nonumber \\
& - & Z_{\beta,0}{\beta \over 2} {\partial \over \partial t_1}
\left \langle \varphi_i(t) \varphi_{j_1}(t_1) P_I(\varphi(t_w)) e^{\beta  {\cal H}_0(\varphi(t_w))}
\right \rangle_{\beta,0} \nonumber \\
& - &  Z_{\beta,0}{\beta \over 2} \left \langle \varphi_i(t) B_{j_1}(t_1) P_I(\varphi(t_w))
e^{\beta  {\cal H}_0(\varphi(t_w))} \right \rangle_{\beta,0}.
\label{Ub.32}
\end{eqnarray}
The next step consists in the recognition that for an arbitrary observable $A$
\be
Z_{\beta,0} \left \langle A(t) P_I(\varphi(t_w))e^{\beta  {\cal H}_0(\varphi(t_w))} \right \rangle_{\beta,0}
= \langle  A(t) \rangle_{I \rightarrow \beta,0}
\label{Ub.33}
\ee
since the factor $Z_{\beta,0}P_I(\varphi(t_w))e^{\beta  {\cal H}_0(\varphi(t_w))}$ in the left hand side
has the effect of undoing the equilibrium initial condition and replacing it with $P_I$.
Hence, Eq.~(\ref{Ub.32}) can be rewritten as
\begin{eqnarray}
& & Z_{\beta,0}
\left . {\delta
\left \langle P_I(\varphi(t_F)) e^{\beta  {\cal H}_0(\varphi(t_F))}  \varphi_i(t_0)
\right \rangle_{\beta,0 \rightarrow \beta,[\widetilde{h}(t)]}  \over
\delta \widetilde{h}_{j_1}(\widetilde{t}_1)} \right |_{\widetilde{h}=0} = \nonumber \\
& - & {\beta \over 2} {\partial \over \partial t_1}
\left \langle \varphi_i(t) \varphi_{j_1}(t_1) \right \rangle_{I \rightarrow \beta,0}
 -   {\beta \over 2} \left \langle \varphi_i(t) B_{j_1}(t_1) \right \rangle_{I \rightarrow \beta,0} \nonumber \\
\label{Ub.34}
\end{eqnarray}
and inserting it into Eq.~(\ref{Ub.30}), eventually one finds
\begin{eqnarray}
& & R^{(1,1)}_{i;j_1}(t,t_1) =
{\beta \over 2} {\partial \over \partial t_1}
\langle \varphi_i(t) \varphi_{j_1}(t_1)\rangle_{I \rightarrow \beta,0} \nonumber \\
& - &  {\beta \over 2} \left \langle \varphi_i(t) B_{j_1}(t_1) \right \rangle_{I \rightarrow \beta,0}
\label{Ub.35}
\end{eqnarray}
thus recovering Eq.~(\ref{RF14}).

\section{Nonlinear susceptibility and growing length scale}
\label{length}

In this section we consider the use of the response functions
in the diagnostics of cooperative effects taking place over large length scales during the
relaxation, referring to the case of Ising spin systems. A partial and preliminary
account of the material in this section has been presented in Ref.~\cite{LCSZ}.

In non disordered coarsening systems, such as a ferromagnet quenched to or to below
the critical point, the existence of a growing dynamical correlation length
is well captured through the scaling properties of the equal time correlation function 
$C_{ij}(t) = M_{ij}(t,t)$~\cite{Bray}, recalling that
in this kind of processes $M_i(t) \equiv 0$. 
In glassy systems, instead,
quenched or self-induced disorder renders the two body correlation function short ranged,
making it necessary to resort to higher order correlation functions. Attention
has been particularly focussed on the four-point correlation function~\cite{Franz}
\be
C^{(4)}_{ij}(t,t_w) = M_{iijj}(t,t_w,t,t_w) -  M_{ii}(t,t_w) M_{jj}(t,t_w)
\label{CCF.1}
\ee
which describes the so called heterogeneities~\cite{hetero}, namely the space fluctuations 
associated to the local time
decorrelation. Although quite convenient in the numerical simulations,
$C^{(4)}_{ij}(t,t_w)$ has the shortcoming of being hardly accessible
in the experiments. Conversely, susceptibilities are more easily measurable
and this has prompted the investigation of the nonlinear FDR.

Here, we expand on the proposals~\cite{Huse,biroli,science,LCSZ} to investigate dynamic scaling through
higher order susceptibilities. Let us begin by considering 
the second order response of the second moment at equal times 
\be
R^{(2,2)}_{ij;j_1j_2}(t,t;t_1,t_2) =  \left .{\delta^2 M_{ij}(t,t,[h_i(t^{\prime}]) \over
\delta h_{j_1}(t_1)\delta h_{j_2}(t_2)}\right |_{h=0}.
\label{CCF.1bis}
\ee
Proceeding exactly as in the derivation of Eq.~(\ref{NL8bis}), the corresponding FDR
is given by
\begin{eqnarray}
R^{(2,2)}_{ij;j_1j_2}(t,t;t_1,t_2)& = & 
( \beta /2 )^2  \Big \{ {\partial \over \partial t_M}{\partial \over \partial t_m}
M_{ijj_Mj_m}(t,t,t_M,t_m)  \nonumber \\
& - & {\partial \over \partial t_M}
\langle \hat{\sigma}^z_i (t) \hat{\sigma}^z_j (t) \hat{\sigma}^z_{j_M}(t_M)
\hat{B}_{j_m}(t_m)\rangle  \nonumber \\
& - & {\partial \over \partial t_m}
\langle \hat{\sigma}^z_i (t) \hat{\sigma}^z_j (t) \hat{B}_{j_M}(t_M)
\hat{\sigma}^z_{j_m}(t_m)\rangle  \nonumber \\
& + &  \langle \hat{\sigma}^z_i (t) \hat{\sigma}^z_j (t) \hat{B}_{j_M}(t_M)
\hat{B}_{j_m}(t_m)\rangle  \Big \} \nonumber \\
& + & (\beta^2/2) \Big \{ \langle \hat{\sigma}^z_i (t) \hat{\sigma}^z_j (t)
\hat{\sigma}^z_{j_m}(t_M)\hat{B}_{j_m}(t_m) \rangle \nonumber \\
& + &  {\partial \over \partial t_m} M_{ijj_mj_m}(t,t,t_M,t_m) \Big \}
\delta_{j_M,j_m} \delta(t_M - t_m). \nonumber \\
\label{CCF.1tris}
\end{eqnarray}
Looking at the above equation, it may seem farfetched to relate the properties of $C^{(4)}$ to
those of $R^{(2,2)}$, since the fourth order moment appears explicitly 
only in the first term in the right hand side and in the singular term.
Nonetheless, information on the existence of a growing correlation length
can be obtained through a scaling argument, as it will be shown below.
Let us consider the more manageable
integrated response function
\be
-\chi^{(2,2)}_{ij}(t,t_w)
 = \int_{t_w}^t dt_1 \int_{t_w}^t dt_2 R^{(2,2)}_{ij;ij}(t,t;t_1,t_2)
- \chi^{(1,1)}_{i}(t,t_w)\chi^{(1,1)}_{j}(t,t_w)
\label{CCF.2}
\ee
where the $\chi^{(1,1)}_i$ in the subtraction are the time integrals of the linear response 
function~(\ref{RF14}), i.e.
\be
\chi^{(1,1)}_{i}(t,t_w) = \int_{t_w}^t dt_1 R^{(1,1)}_{i;i}(t,t_1).
\label{CCF.4}
\ee
The reason for this subtraction and for the minus sign on the left hand side 
will be clear shortly. Assuming that eventually an equilibrium state is reached,
from equilibrium statistical mechanics follows
\be
T^2 \lim _{t\to \infty}\chi^{(2,2)}_{ij}(t,t_w)=C^2_{ij,eq} 
\label{CCF.7}
\ee 
where, for simplicity, we have taken $\langle \hat{\sigma}^z_i \rangle_{eq}=0$.
The scaling relation
$ C_{ij,eq} = \xi^{2-d-\eta} F_{C,eq}  (|i-j|/ \xi )$, where $\xi (T)$ is the equilibrium
correlation length, suggests a finite time scaling behavior of the form
\be
T^2\chi^{(2,2)}_{ij}(t,t_w) = \xi^x F \left ({|i-j| \over \xi},{t^{1/z} \over \xi},{t_w\over t} \right )
\label{CCF.5}
\ee
where $x=4-2d-2\eta$ and $z$ is the dynamical exponent.

Another quantity, which has been recently~\cite{biroli} considered in relation to $C^{(4)}$, 
is the third order integrated response of the first moment
\be
\chi ^{(1,3)}_{ij}(t,t_w)=-\frac{1}{2} \int _{t_w}^t dt_1 dt_2 dt_3 
R_{i;ijj}^{(1,3)}(t;t_1,t_2,t_3)
\label{unmez}
\ee
where $R_{i;ijj}^{(1,3)}$ is given by Eq.(\ref{threekicks}) of Appendix I.
Again, from the large time result
\be
T^3 \lim _{t\to \infty }\chi^{(1,3)}_{ij}(t,t_w)= C^2_{ij,eq} 
\label{CCF.777}
\ee
we may infer the scaling behavior
\be
T^3\chi^{(1,3)}_{ij}(t,t_w) = \xi^x G \left ({|i-j| \over \xi},{t^{1/z} \over \xi},{t_w\over t} \right ).
\label{chi3scal}
\ee
Notice that the prefactor in the definition~(\ref{unmez}) has been arranged in such a way
that the large time limits of $T^2 \chi ^{(2,2)}$ and $T^3\chi ^{(1,3)}$ are the same. 

These patterns of behavior have been checked numerically in the
one-dimensional Ising model. The simulation has been carried out through standard Montecarlo techniques
with Glauber transition rates, where 
$\hat{B}_i=\hat{\sigma}^z_i-\tanh(\beta \sum _{<j>_i} J_{ij}\hat{\sigma}^z_j)$ and the sum runs over the
nearest neighbours $<j>_i$ of $i$.
Taking $J_{ij}=1$, we have prepared the system in a high temperature uncorrelated state
and then quenched it to the final temperature $T$ at time $t=0$.
In this case $z=2$~\cite{Bray}, and $\xi (T)= -1/\ln [\tanh (1/T)]$~\cite{Glauber}.
The integrated response functions $\chi^{(2,2)}_{ij}(t,0)$ and $\chi^{(1,3)}_{ij}(t,0)$ have been computed 
using the FDR~(\ref{CCF.1tris},\ref{threekicks}), following the zero field method of
Ref.~\cite{lippiello05}.
It must be stressed that, due to the extremely noisy nature of the response functions,
the numerical computation of these quantities through the FDR is by far more convenient
than the computation based on the application of
a small external field. In fact,
in addition to the incomparably better signal to noise ratio,
the $h\to 0$ limit is built in the FDR.

\begin{figure}
    \centering

   \rotatebox{0}{\resizebox{.7\textwidth}{!}{\includegraphics{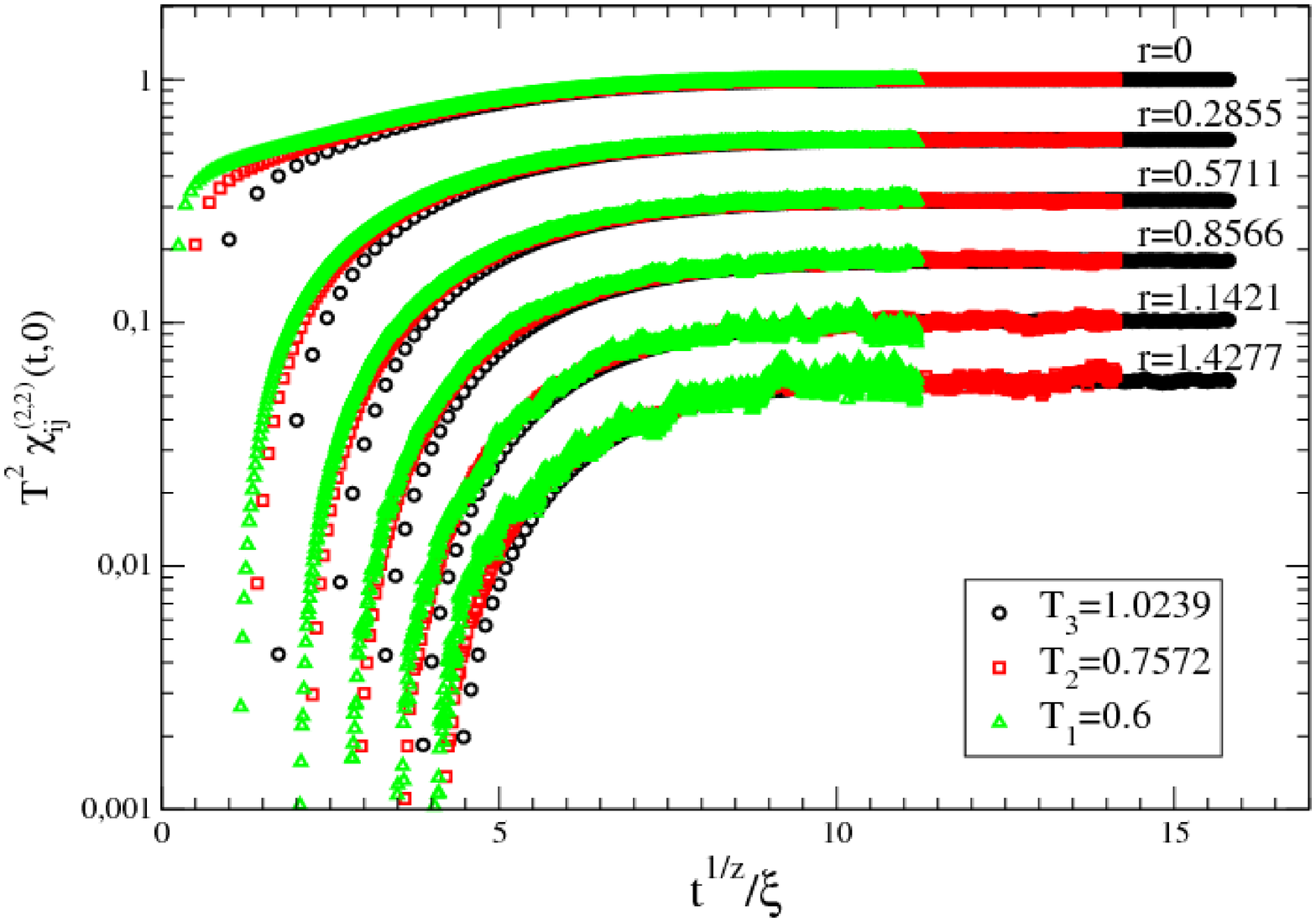}}}

   \vspace{1.7cm}

   \rotatebox{0}{\resizebox{.7\textwidth}{!}{\includegraphics{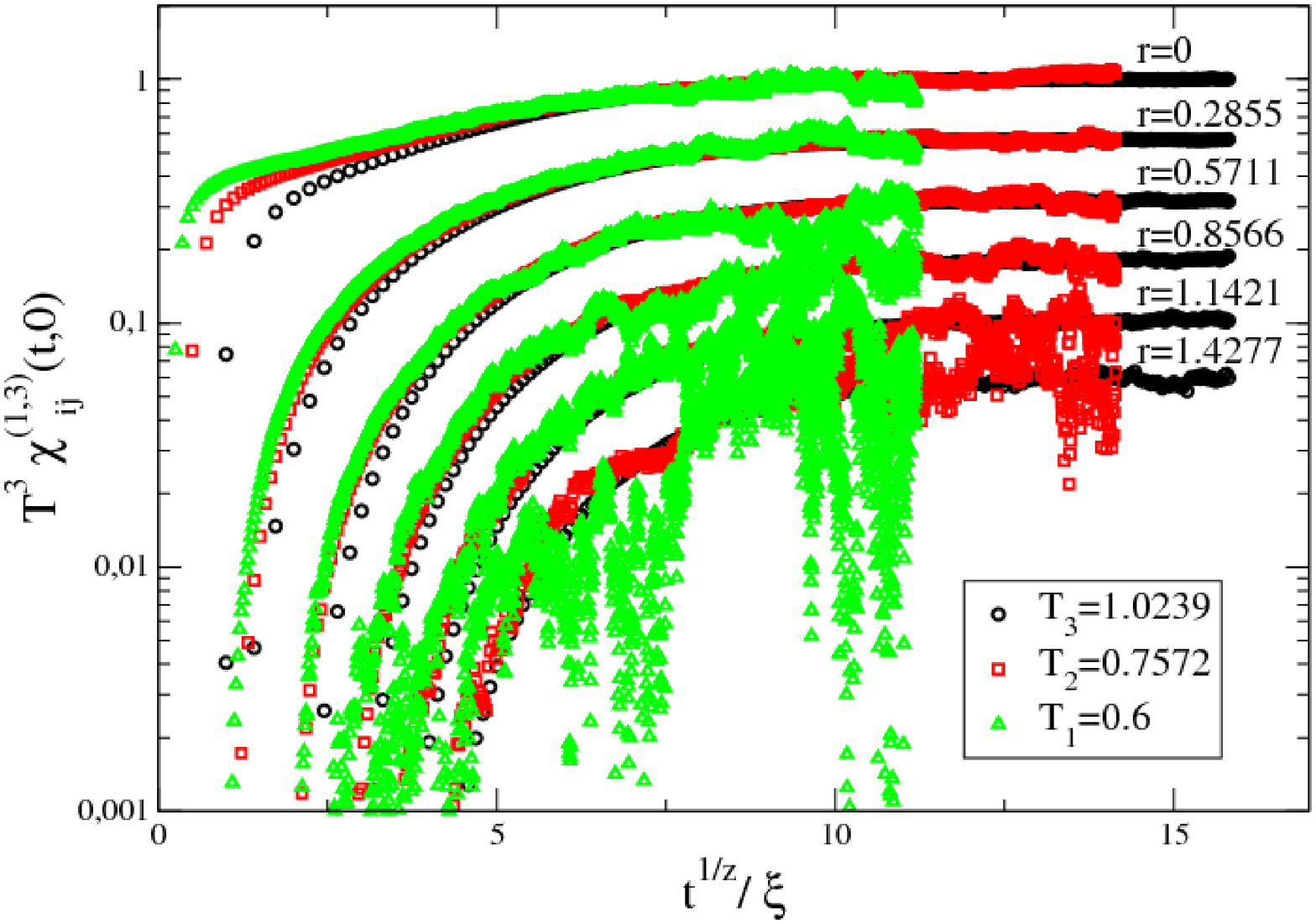}}}

   \vspace{1cm}
   
    \caption{(Color online). Upper panel: The quantity $T^2 \chi^{(2,2)}_{ij}(t,0)$ 
is plotted against 
$t^{1/z}/\xi$ ($z=2$) for six fixed values of 
$r=|i-j|/\xi$ ($r=0, 0.2855, 0.5711, 0.8566, 1.1421, 1.4277$
from top to bottom) and  
three temperatures  $T_1=0.6, T_2=0.7572, T_3=1.0239$ such
that $\xi (T_1)/\xi (T_2)=2$ and  $\xi (T_1)/\xi (T_3)=4$.
Lower panel: Same for $T^3 \chi ^{(1,3)}_{ij}(t,0)$.}  
\label{figurachi23}
\end{figure}

In order to verify the scaling relations~(\ref{CCF.5}) and~(\ref{chi3scal}), first of all we have taken $t_w=0$
in order to reduce the variables from three to two in the right hand sides. 
Then, recalling that for the $1d$ Ising model $\eta =1$ implies $x=0$ and
varying the temperature and the distance in such a way to keep
$|i-j|/\xi$ fixed, it is matter of showing that $T^2\chi^{(2,2)}_{ij}(t,0)$ and $T^3\chi^{(1,3)}_{ij}(t,0)$ are 
functions only of $t^{1/z}/\xi$. 
This is shown, with good accuracy, in Fig.~(\ref{figurachi23}) for the quenches to the three
final temperatures $T_1=0.6, T_2=0.7572$ and $T_3=1.0239$ and with
six different values of $|i-j|/\xi$. 
Both $T^2\chi^{(2,2)}_{ij}(t,0)$ and $T^3\chi^{(1,3)}_{ij}(t,0)$  grow from
zero to the same asymptotic value $C_{ij,eq} ^2=\exp \{-2|i-j|/\xi \}$
on the same timescale, as expected.
Note that the data for
$\chi^{(1,3)}$ are much more noisy than those for $\chi^{(2,2)}$.
Since the same numerical resources have been allocated in the 
computation of each of these two quantities,
we conclude that investigations based on $\chi^{(2,2)}$ are more 
efficient, at least numerically.

We now discuss how $\chi^{(2,2)}$ can be effectively used for the measurement
of a cooperative length in disordered systems.
In order to improve the statistics further, it is convenient
to consider the $\vec{k}=0$ component of the space Fourier transform
\be
\chi^{(2,2)}_{\vec{k}=0}(t,t_w)=(1/N^2)\sum _{i,j}\chi^{(2,2)}_{ij}(t,t_w)
\label{88}
\ee 
which, using Eq.~(\ref{CCF.5}), scales as
\be
\chi^{(2,2)}_{\vec{k}=0}(t,t_w) = \xi^{4-d-2\eta}
{\cal F} \left ({t^{1/z} \over \xi},{t_w\over t} \right ).
\label{CCF.8}
\ee

\begin{figure}
    \centering

   \rotatebox{0}{\resizebox{.7\textwidth}{!}{\includegraphics{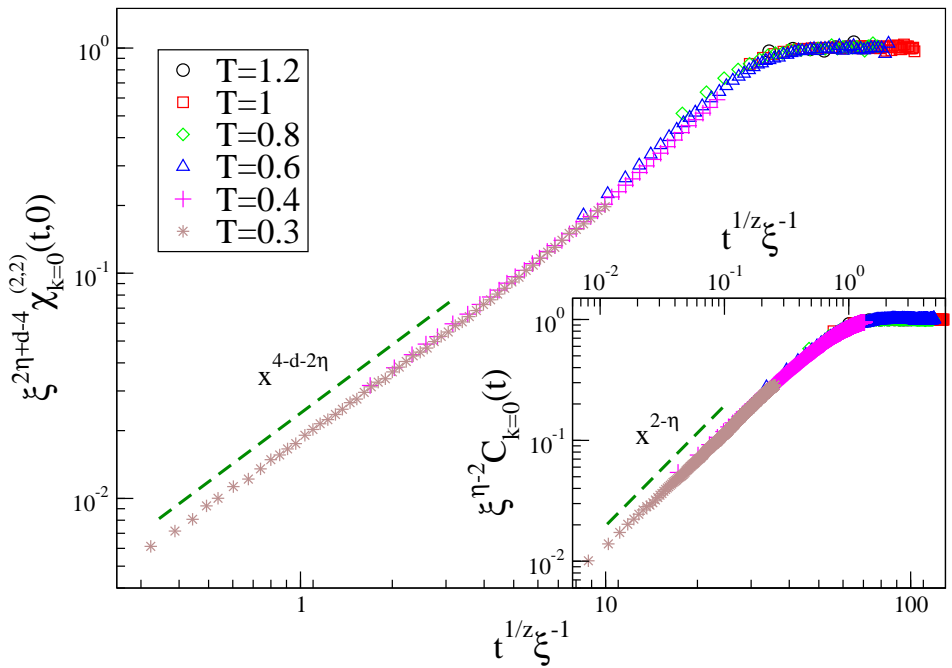}}}
\vspace{1cm}
    \caption{(Color online). Data collapse of $\chi ^{(2,2)}_{k=0}$ ($C_{k=0}$ in the inset) 
for several temperatures
in the $d=1$ EA model ($\eta=1, z=2$). The dashed lines are the expected power-laws in the non 
equilibrium regime.}

\label{d1}
\end{figure}

\begin{figure}
    \centering

   \rotatebox{0}{\resizebox{.7\textwidth}{!}{\includegraphics{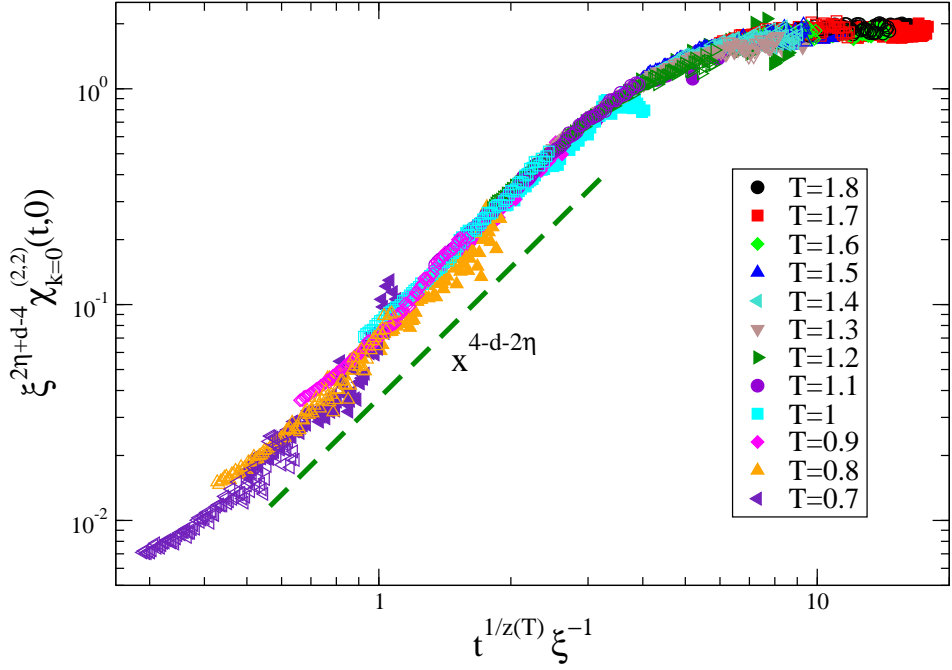}}}
    \caption{(Color online). Data collapse of $\chi ^{(2,2)}_{k=0}$
for several $T$ in the $d=2$ EA model with
bimodal (open symbols) or Gaussian (filled symbols) bond distribution,
with $z(T)=4/T$ and $\eta =0$.
The dashed line is the expected power-law in the non equilibrium regime.}
\label{d2}
\end{figure}

In Ref.~\cite{LCSZ} we have computed numerically this quantity, with $t_w=0$, in the
Edwards-Anderson (EA) model with Hamiltonian $\hat{{\cal H}}=\sum _{ij} J_{ij} \hat{\sigma}^z_i \hat{\sigma}^z_j$
and $d=1,2$. 
The data have been analysed as follows: 
for large $t$ the curves saturate to the equilibrium value 
$\chi^{(2,2)}_{\vec{k}=0,eq} \sim \xi^{4-d-2\eta}$.
Using the known values of $\eta $,
one can extract $\xi$.
In the off-equilibrium regime $ t^{1/z} \ll \xi$ the growing correlation length
$L(t)\sim t^{1/z}$ is expected not to depend on $\xi$.
Enforcing this condition 
from Eq.~(\ref{CCF.8}) one must have ${\cal F} (t^{1/z} / \xi ,0) \sim
(\xi/L(t))^{2\eta+d-4}$, which yields
\be
\chi^{(2,2)}_{\vec{k}=0}(t,0) \sim L(t)^{4-d-2\eta}.
\label{CCF.9}
\ee
This allows to determine $L(t)$.
After doing this, we checked for the data collapse by plotting
$\xi^{-4+d+2\eta}\chi ^{(2,2)}_{k=0}(t,0)$ vs $L(t)/\xi$ for all the temperatures
considered (see Figs.~\ref{d1},\ref{d2}).

Let us consider first
the  $d=1$  EA model with bimodal distribution of the coupling constants $J_{ij}=\pm 1$.
This system is considered in order to test the method, since
it can be mapped onto the ferromagnetic system, just considered above. 
Moreover, in this simple case, in addition to  $\chi ^{(2,2)}_{k=0}(t,0)$,
one can check the scaling
of the equal time structure factor $C_{k=0}(t)$, obtaining
independent determinations of $L(t)$ and $\xi $ to compare with.
This  shows that the sets of data for $L(t)$ and $\xi$, obtained in both ways,
are in agreement with each other and with the analytical behaviors
up to the numerical uncertainty.
The data collapse of $\chi ^{(2,2)}_{k=0}(t,0)$ and of $C_{k=0}(t)$ are shown in Fig.~\ref{d1}.
Here, one clearly observes the off equilibrium regime,
characterized by the power-law behavior of $\chi ^{(2,2)}_{k=0}(t,0)$
and $C_{k=0}(t)$
with exponents $4-d-2\eta$ and $2-\eta $, respectivly.
In the large time regime 
equilibration takes place with the convergence of
$\chi ^{(2,2)}_{k=0}(t,0)$ and $C_{k=0}(t)$ to $\chi ^{(2,2)}_{k=0,eq}$ and to $C_{k=0,eq}(t)$.

After this explicit verification, we have turned to the $d=2$ case,
where the independent information on the structure factor is not available.
Both with bimodal and Gaussian distributions of $J_{ij}$,
the behavior of $\xi$ extracted from $\chi^{(2,2)}_{\vec{k}=0,eq}$,
using $\eta =0$~\cite{sgd2bis}, has been found
consistent with previous results~\cite{sgd2bis,sgd2}.
The non-equilibrium behavior is compatible with a power law
$L(t)\sim t^{1/z(T)}$ with a temperature dependent exponent in agreement
with $z(T)\simeq 4/T$, as reported in Ref.~\cite{sgd2z}.
The data collapse of $\chi ^{(2,2)}_{k=0}(t,0)$ is shown in Fig.~\ref{d2}. Notice also the
additional collapse of the curves with bimodal and Gaussian bond distribution,
further suggesting that the two models may share the same universality class
at finite temperatures~\cite{sgd2bis}.

\section{Effective temperature}

One of the most interesting developments in the study of the linear
FDR out of equilibrium has been the introduction of the concept of
effective temperature $T_{eff}$~\cite{Peliti}. The idea is that the off equilibrium
behavior observed during slow relaxation can be accounted for
by the separation of the time scales for different subsets of degrees
of freedom. Each of these is regarded as in equilibrium with a different virtual thermostat
at some appropriate effective temperature, which depends on the time scale
and is different from the physical temperature of the real
thermostat driving the relaxation. The value of $T_{eff}$ can be inferred
by forcing the off equilibrium linear FDR in the form of the equilibrium
FDT. Although appealing, this idea has turned out not to
be applicable {\it tout court}, since $T_{eff}$ might turn out to be observable
dependent~\cite{Gambassi}. Nonetheless, with the proper caveats, the concept remains
quite useful and suggestive. In this section we make a preliminary exploration of another
open end in the important question of how general $T_{eff}$ can be,
investigating whether it is possible to extend to
the nonlinear FDR the effective temperature concept, consistently with what it is done
in the linear case.

Let us first recall how $T_{eff}$ is defined from the linear FDR.
For definiteness, the Ising spin case will be considered. Writing explicitely
the time integral in Eq.(\ref{CCF.4}), one has 
\begin{eqnarray}
\chi^{(1,1)}_{i}(t,t_w) = 
{\beta \over 2} \int_{t_w}^t dt_1 \left [{\partial \over \partial t_1} M_{ii}(t,t_1)
-  \langle \sigma^z_i (t) \hat{B}_{i}(t_1) \rangle \right ]. 
\label{eff1}
\end{eqnarray}
Assuming $M_i(t) \equiv 0$ throughout the dynamical evolution, and replacing  $M_{ii}(t,t_1)$
with the autocorrelation function $C(t,t_1)$, 
the quantity
\be
\psi^{(1)}(t,t_w) =  \int_{t_w}^t dt_1 {\partial \over \partial t_1}C(t,t_1)  = 1- C(t,t_w)
\label{eff3}
\ee
for fixed $t_w$ is a monotonously increasing function of time, which allows to reparametrize
$t$ in terms of $\psi^{(1)}$ and to write $\chi^{(1,1)}_i(t,t_w)$ in the form
\be
\chi^{(1,1)}_i(t,t_w) = \chi^{(1,1)}_i(\psi^{(1)},t_w).
\label{eff5}
\ee
In equilibrium, where time translation invariance holds, the dependence on $t_w$ disappears and the parametric 
representation becomes linear
\be
\chi^{(1,1)}_i(\psi^{(1)}) = \beta \psi^{(1)} 
\label{eff6}
\ee
with the obvious consequence
\be
\beta = {d\chi^{(1,1)}_i(\psi^{(1)}) \over d\psi^{(1)}}.
\label{eff7}
\ee
Off equilibrium, the parametric representation won't be linear and an effective temperature
can be defined by the generalization of the above relation
\be
\beta_{eff}(\psi^{(1)},t_w) = { \partial \chi^{(1,1)}_i(\psi^{(1)},t_w)
\over \partial\psi^{(1)}}
\label{eff8}
\ee 
with $\beta_{eff} = 1/T_{eff}$.

In order to see how $T_{eff} \neq T$ arises in a simple context, let us consider
the relaxation to a low temperature phase characterized by
ergodicity breaking and, therefore, by a non vanishing Edwards-Anderson
order parameter $q_{EA}$. In particular, let us think of the already mentioned
coarsening process, like in a ferromagnet quenched to below the critical point and relaxing via domain growth.
In that case $q_{EA}$ coincides with the spontaneous magnetization squared $M_{eq}^2$. 

\begin{figure}
    \centering

   \rotatebox{0}{\resizebox{.8\textwidth}{!}{\includegraphics{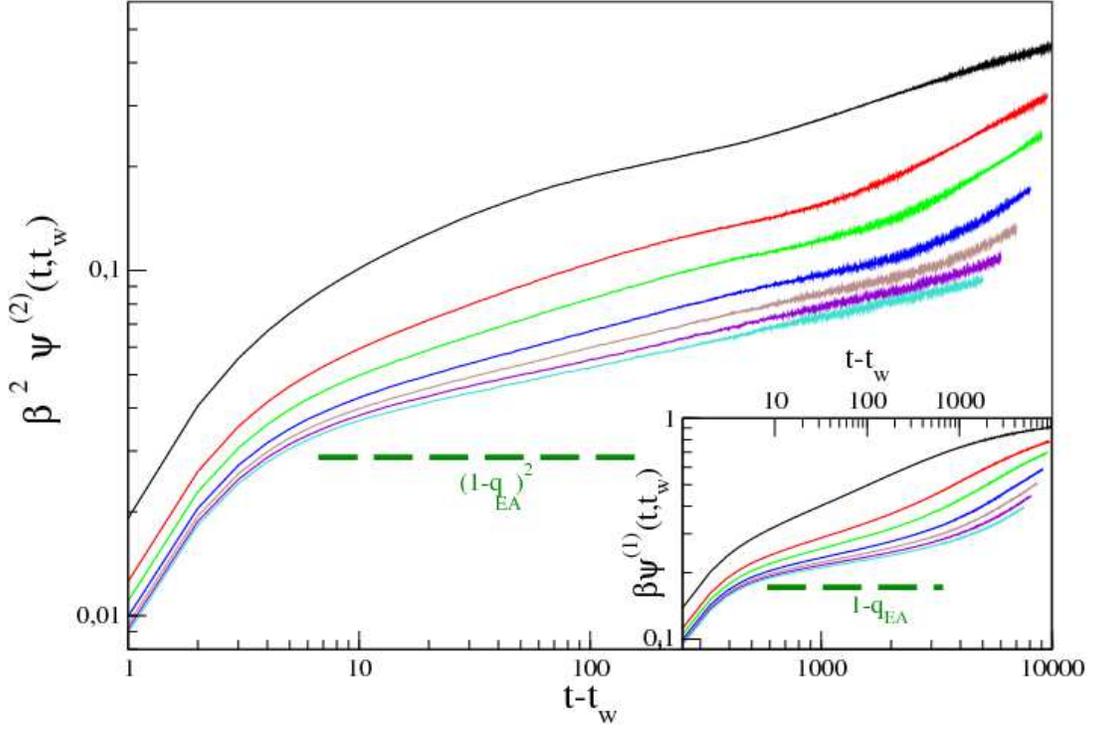}}}
\vspace{1cm}
    \caption{(Color online). The quantity $\psi ^{(2)}$ is plotted against $t-t_w$
for $t_w=100,500,1000,2000,3000,4000,5000$ from top to bottom. Data refer to
a $2d$ Ising system of size $1800^2$ quenched to $T=2$ ($T_c\simeq 2.269$). The 
dashed horizontal line is the large-$t_w$ height of the plateau ($y=(1-q_{EA})^2$).
In the inset the quantity
$\psi ^{(1)}$ is shown (same $t_w$ of the main figure).} 
\label{figurapsi}
\end{figure}

As $t_w \rightarrow \infty$, the  separation of time scales takes place.
The short, or quasiequilibrium, time regime holds for $C> M_{eq}^2$, that is $\psi^{(1)} < 1- M_{eq}^2$,
while the large time scale sets in
when $C < M_{eq}^2$, or $\psi^{(1)} > 1- M_{eq}^2$. The existence of this latter regime
makes it evident the failure of equilibration,
even in the $t_w \rightarrow \infty$ limit, since the autocorelation function falls below
the Edwards-Anderson plateau.
The behavior of $\psi ^{(1)}$, obtained from numerical simulations of the Ising model in $d=2$, 
is shown in the inset of Fig.~\ref{figurapsi}.
Starting from zero, there is a fast growth in the short time regime, followed by
a plateau, more evident for large $t_w$, and finally there is convergence, 
with the power law behavior $1-\psi ^{(1)}(t,t_w)\sim t^{-\lambda /z }$,
toward the asymptotic value $\psi ^{(1)}=1$. Notice that
$\lambda $ is the Fisher-Huse exponent~\cite{FH} and that 
the plateau flattens over the asymptotic value $1-q_{EA}$ as $t_w \rightarrow \infty$.

Correspondingly, the integrated response function can be written as the sum of two 
pieces~\cite{BCKM}
\be
\chi^{(1,1)}_i(\psi^{(1)},t_w) = \chi^{(1,1)}_{st}(\psi^{(1)})
+ \chi^{(1,1)}_{ag}(\psi^{(1)},t_w)
\label{eff9}
\ee 
where $\chi^{(1,1)}_{st}$ is the stationary contribution arising from the equilibrated
bulk of domains, while $\chi^{(1,1)}_{ag}$ is the aging contribution
due to the off equilibrium domain walls.
The stationary contribution
obeys Eq.~(\ref{eff6}) in the short time regime, saturates to its equilibrium value
and remains constant in the large time regime,
while the aging contribution vanishes as $t_w \rightarrow \infty$ according to
\be
\chi^{(1,1)}_{ag}(\psi^{(1)},t_w) = t_w^{-a}F(\psi^{(1)}) 
\label{eff11}
\ee 
where $a >0$~\cite{noi}. Hence, the full response function obeys the asymptotic form
\be
\lim_{t_w \to \infty}\chi^{(1,1)}_i(\psi^{(1)},t_w) = \left \{ \begin{array}{ll}
        \beta \psi^{(1)}, \;\;$for$ \;\; 0 \leq  \psi^{(1)} \leq 1-M_{eq}^2  \\
        \beta (1-M_{eq}^2),  \;\; $for$ \;\; 1-M_{eq}^2 < \psi^{(1)} \leq 1
        \end{array}
        \right .
        \label{eff10}
        \ee
which, on account of Eq.~(\ref{eff8}), leads to
\be
\lim_{t_w \to \infty}\beta_{eff}(\psi^{(1)},t_w)    = \left \{ \begin{array}{ll}
        \beta, \;\;$for$ \;\; 0 \leq  \psi^{(1)} \leq 1-M_{eq}^2  \\
        0,  \;\; $for$ \;\; 1-M_{eq}^2 < \psi^{(1)} \leq 1.
        \end{array}
        \right .
        \label{eff12}
        \ee
Namely, the effective temperature coincides with the physical temperature in the short time
regime, where the sytem appears equilibrated, while it is drastically different from it in the off
equilibrium large time regime.

Let us now carry out the parallel analysis on the $\vec k=0$ component
of the second order integrated response function
\be
\chi^{(2,2)}_{\vec k=0}(t,t_w) = \frac{1}{N^2}\sum _{i,j}\int_{t_w}^t dt_1 
\int_{t_w}^t dt_2 R^{(2,2)}_{ij;ij}(t,t;t_1,t_2).
\label{eff13}
\ee
Notice that, although we use the same notation, 
this quantity differs from the one in Eq.~(\ref{88}), 
because of the overall sign and of the absence of the subtraction.
Let us introduce the quantity
\begin{eqnarray}
\psi^{(2)}(t,t_w) &=& \frac{1}{2}\frac{1}{N^2}\sum _{i,j}\int_{t_w}^t dt_1 
\int_{t_w}^t dt_2 \Big [ \frac{\partial ^2}{\partial t_M \partial t_m}
M_{ijij}(t,t,t_M,t_m) \nonumber \\
&+& \langle \sigma^z_i(t)\sigma^z_j(t)\hat{B}_i(t_M)\hat{B}_j(t_m)\rangle \Big ]
\label{eff14}
\end{eqnarray}
with properties similar to those of $\psi^{(1)}(t,t_w)$, as shown in
Fig.~\ref{figurapsi}. 
The main feature is the monotonous increase from zero to an asymptotic value
well above the limiting value $(1-q_{EA})^2$ that one would get from the equilibrium
calculation. This is the value at which a plateau develops as $t_w$ gets large,
signaling the separation of time scales. Using $\psi^{(2)}$ to reparametrize the time $t$,
from Eq.~(\ref{TK3}) follows that at equilibrium the FDR becomes
linear
\be
\chi^{(2,2)}_{\vec k=0}(\psi^{(2)}) = {\beta^2 }\psi^{(2)}.
\label{eff15}
\ee
Hence, by following the same reasoning as in the linear case, in the off equilibrium
regime an effective temperature can be introduced by the analogue of Eq.~(\ref{eff8})
\be
\beta^2_{eff}(\psi^{(2)},t_w) = { \partial \chi^{(2,2)}_{\vec k=0}(\psi^{(2)},t_w)
\over \partial\psi^{(2)}}.
\label{eff16}
\ee 
The question, now, is whether the two $\beta_{eff}$ defined by Eqs.~(\ref{eff8}) and~(\ref{eff16})
are consistent or not, that is whether the equality
\be
\lim_{t_w \to \infty}\beta_{eff}(\psi^{(2)},t_w) = \lim_{t_w \to \infty}\beta_{eff}(\psi^{(1)},t_w)
\label{consist}
\ee
holds or not.
This is a difficult question to answer in general, we shall limit 
to the consideration of the particular coarsening process analysed above in the linear case.

The short and the large time scales, in terms of $\psi^{(2)}$, correspond to $\psi^{(2)}$
smaller or larger than $(1-M_{eq}^2)^2$,
respectively. Writing  $\chi^{(2,2)}_{\vec k=0}$ as the sum of two
pieces, as in Eq.~(\ref{eff9}),
\be
\chi^{(2,2)}_{\vec k=0}(\psi^{(2)},t_w) = \chi^{(2,2)}_{st}(\psi^{(2)})
+ \chi^{(2,2)}_{ag}(\psi^{(2)},t_w)
\label{eff17}
\ee 
the same considerations
made on $\chi^{(1,1)}_{st}$
apply exactly to  $\chi^{(2,2)}_{st}$, since this is an equilibrium contribution.
Namely, after obeying Eq.~(\ref{eff15}) in the short time regime, saturates
to the equilibrium value $(1-M_{eq}^2)^2$ and then remains constant in the large time regime.
For $\chi^{(2,2)}_{ag}(\psi^{(2)},t_w)$ there are no previous results to rely on.
We have, then, measured $\chi^{(2,2)}_{ag}(\psi^{(2)},t_w)$ numerically in the
quench of a two dimensional Ising model below $T_C$ evolving with Glauber dynamics. 
The aging contribution  $\chi^{(2,2)}_{ag}$ has been isolated using the
method based on the no-bulk-flip dynamics discussed in~\cite{noi,expab,nbf,commentesteso}.

\begin{figure}
    \centering

   \rotatebox{0}{\resizebox{.55\textwidth}{!}{\includegraphics{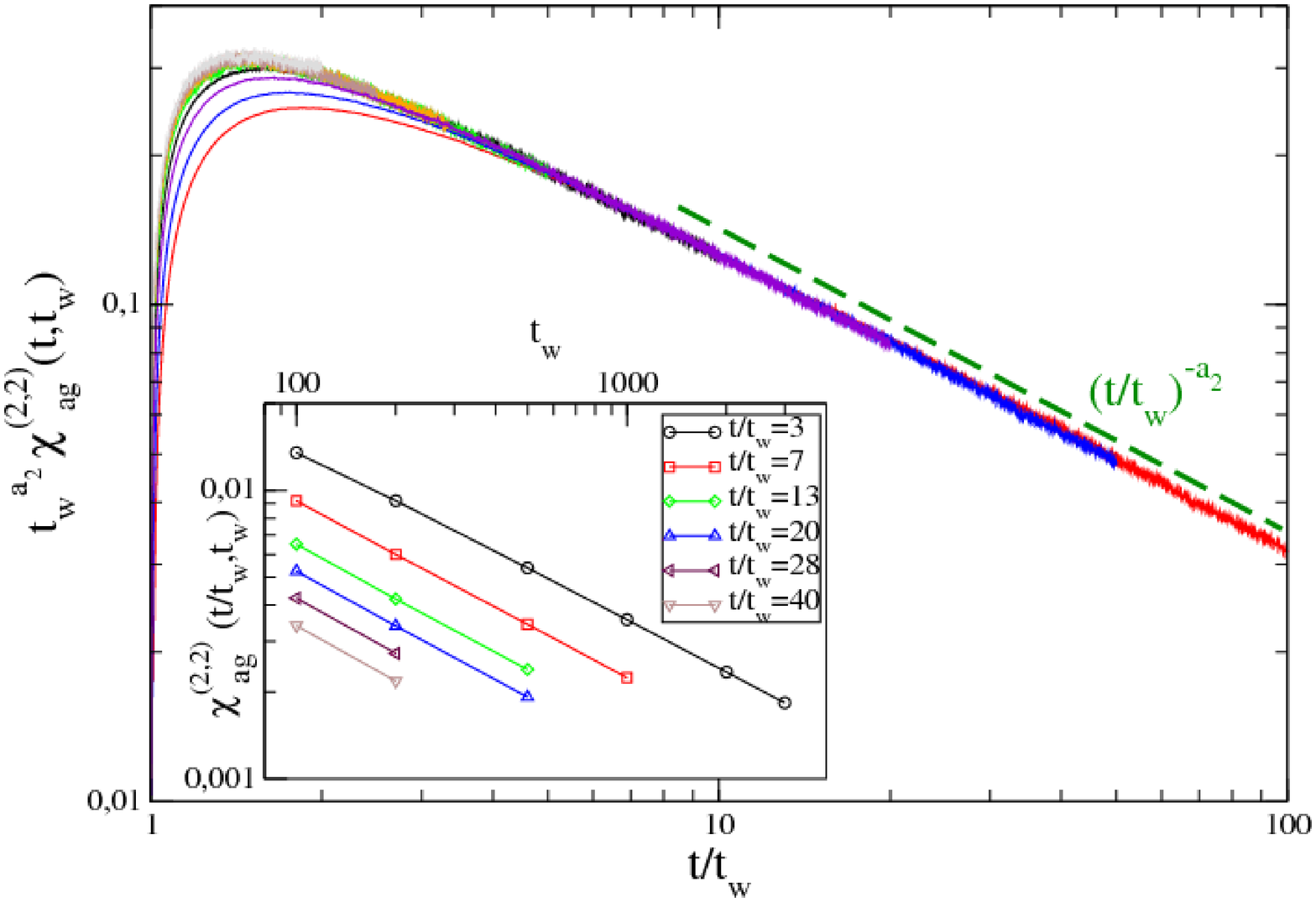}}}

   \vspace{1.2cm}

   \rotatebox{0}{\resizebox{.55\textwidth}{!}{\includegraphics{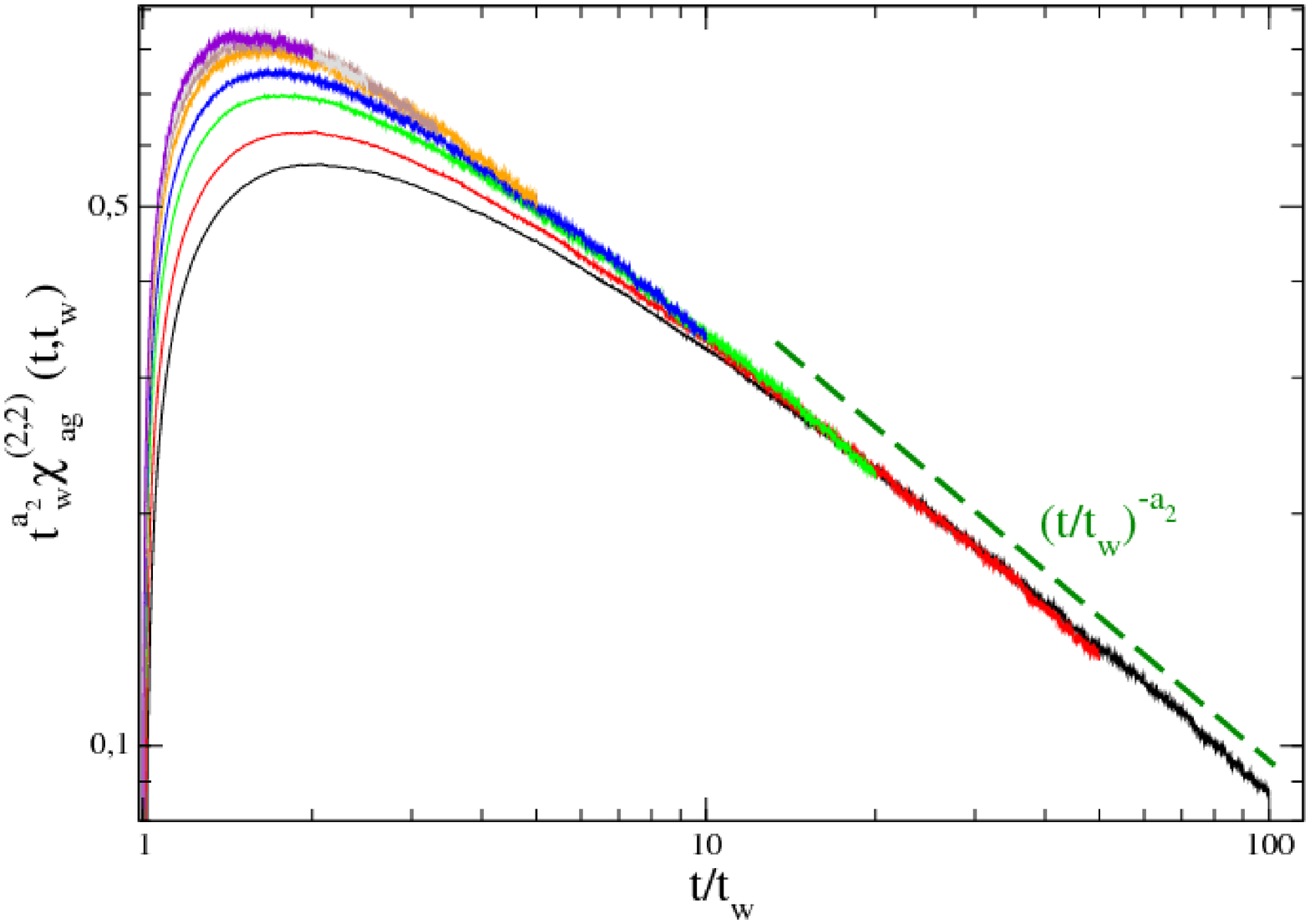}}}

   \vspace{1.2cm}

   \rotatebox{0}{\resizebox{.55\textwidth}{!}{\includegraphics{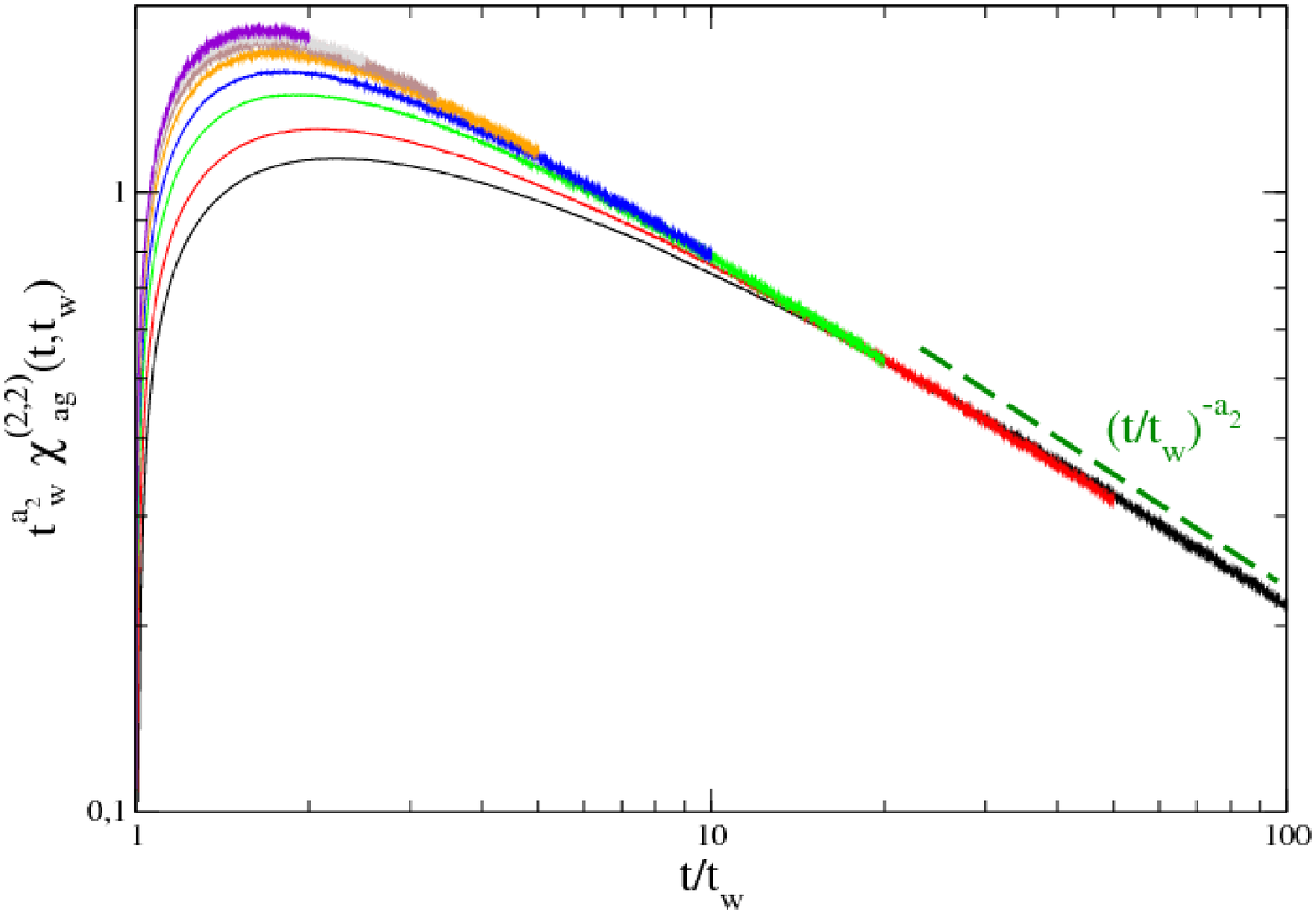}}}
\vspace{1cm}
    \caption{(Color online). $t_w^{a_2}\chi ^{(2,2)}_{ag}(t,t_w)$ is plotted against $t/t_w$
    for $t_w=100,200,500,1000,2000,3000,4000,5000$ (from bottom to top) and
    three temperatures $T=1, 1.5,2$ (upper, central and lower panel). The
    exponent $a_2$ is extracted as discussed in the text, finding $a_2=0.61$
    for $T=1$, and $a_2=0.62$
    for $T=1.5$ and $T=2$. The dashed lines represent the asymptotic behavior
    $\chi^{(2,2)}_{ag}(t,t_w)\sim (t/t_w)^{-a_2}$ for large $t/t_w$.
    In the inset the quantity $\chi ^{(2,2)}_{ag}(t,t_w)$ is plotted against $t_w$
    with $t/t_w$ fixed.}
\label{figuranob}
\end{figure}

In order to check if a scaling form  of the type~(\ref{eff11}) is obeyed
\be
\chi^{(2,2)}_{ag}(\psi^{(2)},t_w) = t_w^{-a_2}F_2(t/t_w)
\label{eff112}
\ee 
we have plotted $\chi^{(2,2)}_{ag}(t,t_w)$ for a fixed value of $t/t_w$ against $t_w$, as shown in the inset
of the upper panel of Fig.~\ref{figuranob}. From the observed power law behavior
we have extracted the exponent $a_2$, finding values in the range $[0.59-0.62]$.
We have, then, carried out the data collapse by plotting 
$t_w ^{a_2}\chi^{(2)}_{ag}(t,t_w)$ against $t/t_w$.
For large $t_w$ the collapse (Fig.~\ref{figuranob}) is quite good,  
confirming that the scaling form~(\ref{eff112}) is obeyed with an exponent $a_2\simeq 0.61-0.62$. 
For small values of $t_w$ and $t/t_w$ deviations are observed due to preasymptotic effects,
similarly to what was already observed in the linear case~\cite{noi,commentesteso}.
Notice, also, that
for large $t/t_w$ one has a power law decay of the scaling function
$F_2(t/t_w)\sim (t/t_w)^{-a_2}$ with the same exponent
$a_2$ entering Eq.~(\ref{eff112}), exactly as it was observed in the linear case~\cite{commentesteso}.

\begin{figure}
    \centering

   \rotatebox{0}{\resizebox{.8\textwidth}{!}{\includegraphics{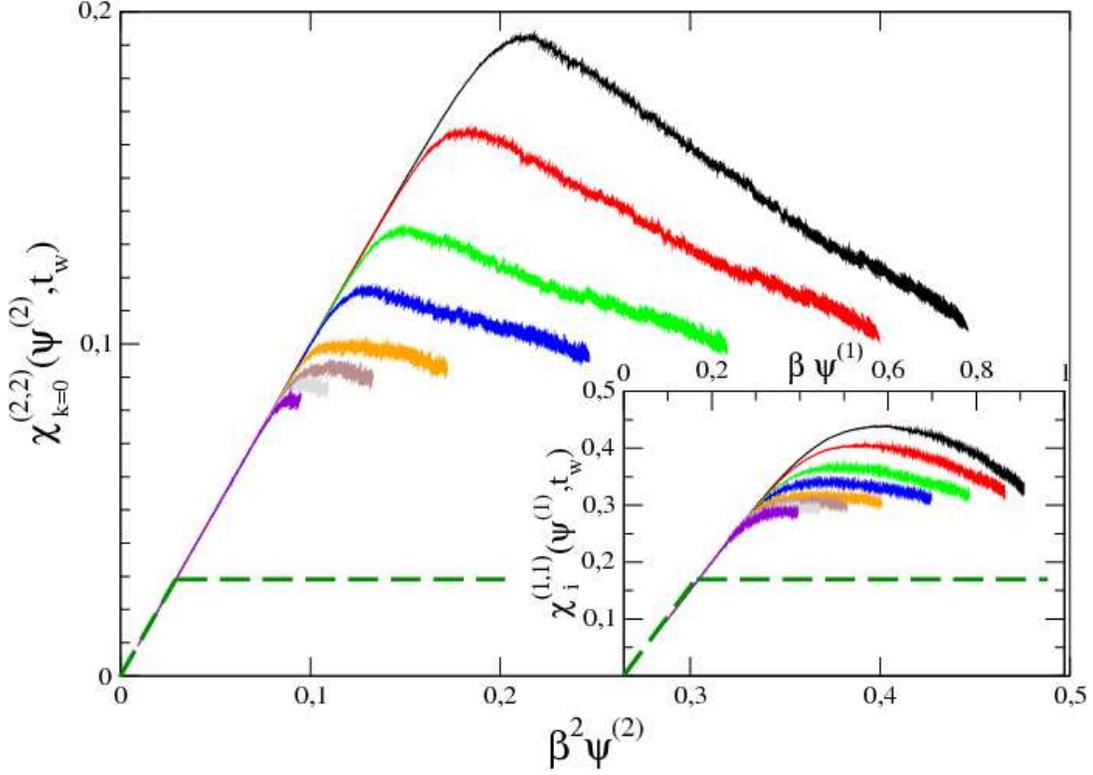}}}
\vspace{1cm}
    \caption{(Color online). The parametric plot of 
    $\chi ^{(2,2)}_{k=0}(t,t_w)$ against $\beta ^2\psi ^{(2)}$ ($\chi ^{(1,1)}_i(t,t_w)$ 
    against $\beta \psi ^{(1)}$ in the inset) is shown for $T=2$ and $t_w=100,200,500,1000,2000,
    3000,4000,5000$, from top to bottom. The dashed line is the expected asymptotic
    behavior.} 
\label{figuraplotpara}
\end{figure}

In conclusion, like in the linear case,
the existence of the scaling behavior~(\ref{eff112}) with $a_2 > 0$ implies that 
the aging contribution vanishes  asymptotically, yielding the analogue 
of Eq.~(\ref{eff10})
\be
\lim_{t_w \to \infty}\chi^{(2,2)}_{\vec{k}=0}(\psi^{(2)},t_w) = \left \{ \begin{array}{ll}
        \beta^2 \psi^{(2)}, \;\;$for$ \;\; \psi^{(2)} \leq (1-M_{eq}^2)^2  \\
        \beta ^2 (1-M_{eq}^2)^2,  \;\; $for$ \;\; (1-M_{eq}^2)^2  < \psi^{(2)}. 
        \end{array}
        \right .
        \label{eff18}
        \ee
The approach to this asymptotic behavior is shown in Fig.~\ref{figuraplotpara}. Hence,
using the definition~(\ref{eff16}), we find
\be
\lim_{t_w \to \infty}\beta_{eff}(\psi^{(2)},t_w)    = \left \{ \begin{array}{ll}
        \beta, \;\;$for$ \;\; \psi^{(2)} \leq (1-M_{eq}^2)^2   \\
        0,  \;\; $for$ \;\; (1-M_{eq}^2)^2  < \psi^{(2)}.
        \end{array}
        \right .
        \label{eff19}
\ee
The comparison with Eq.~(\ref{eff12}) suggests that
the consistency condition~(\ref{consist}) is satisfied.

\section{Conclusions}

In this paper we have derived the off equilibrium FDR of arbitrary order
for systems evolving with Markovian stochastic dynamics. The main
effort has been to put the FDR in the same form for both
continous and discrete spins. In order to stress the generality of
the result, we have also shown how the whole
hierarchy of FDR can be made to descend from the fluctuation principle.
Once the FDR are available, response functions of arbitrary order are
expressed in terms of unperturbed correlation functions of observables.
The payoff is in  the development of simple and efficient zero field algorithms
for the numerical simulations.  

As an application, we have considered the problem of detecting
the existence of a growing length in those cases, like in glassy
systems, where standard methods based on two-point correlation functions
and the corresponding linear response functions are of no use. 
In these cases the simplest object carrying useful information,
in principle, would be a four-point correlation function which,
however, is not directly accessible to experiment. Instead, experimentally
accessible are the nonlinear response functions involving
the four-point correlation function through the nonlinear FDR.
The choice of which response function and, therefore, of which FDR to
use is not univocal, once the realm of the nonlinear response functions 
is entered. Then the choice is matter of convenience.
We have made the proposal to use the second order response
of a two-point correlation function, rather than the third order
response of the magnetization, as advocated elsewhere in the literature.
We have, then, demonstrated the numerical advantage of our choice through
the implementation of the zero field algorithm.

Finally, we have made a first step into the
important but difficult problem of the definition of the effective
temperature through the nonlinear FDR. We have considered
the domain coarsening process ensuing the quench of a ferromagnet
below its critical point. Indeed, in that case we have found 
that it is possible to extract from the nonlinear FDR an
effective temperature which is consistent with the effective temperature
obtained from the much studied linear FDR.

\section{Appendix I}

Proceeding like in the derivation of Eq.~(\ref{respf.7}), the third order derivative
of the propagator is given by
\begin{eqnarray}
& &{\delta^3 \hat{P}_h(t| t_w) \over \delta h_{j_1}(t_1) \delta h_{j_2}(t_2) \delta h_{j_3}(t_3) }  = \nonumber \\
& & \hat{P}_h(t| t_M){\partial \hat{W}(t_M) \over \partial h_{j_M}(t_M)} \hat{P}_h(t_M| t_I)
{\partial \hat{W}(t_I) \over \partial h_{j_I}(t_I)}\hat{P}_h(t_I| t_m)
{\partial \hat{W}(t_m) \over \partial h_{j_m}(t_m)}\hat{P}_h(t_m| t_w) \nonumber \\
& + & \hat{P}_h(t| t_M){\partial^2 \hat{W}(t_I) \over \partial h^2_{j_I}(t_I)} \hat{P}_h(t_I| t_m)
{\partial \hat{W}(t_m) \over \partial h_{j_m}(t_m)}\hat{P}_h(t_m| t_w) \delta_{j_M,j_I}\delta(t_M-t_I) \nonumber \\
& + & \hat{P}_h(t| t_M){\partial\hat{W}(t_M) \over \partial h_{j_M}(t_M)} \hat{P}_h(t_M| t_I)
{\partial^2 \hat{W}(t_I) \over \partial h^2_{j_I}(t_I)}\hat{P}_h(t_I| t_w) \delta_{j_I,j_m}\delta(t_I-t_m) \nonumber \\
& + & \hat{P}_h(t| t_1){\partial^3 \hat{W}(t_1) \over \partial h^3_{j_1}(t_1)} \hat{P}_h(t_1| t_w)
\delta(12)\delta(23)
\label{RF.02bis}
\end{eqnarray}
where $t_M = \max(t_j)$, $t_m = \min(t_j)$, $t_m \leq t_{I} \leq t_M$,
$j_M,j_I,j_m$ are the sites where the field acts at the times $t_M$, $t_I$ or $t_m$, respectively,
and $\delta(np) = \delta_{j_n,j_p}\delta(t_n-t_p)$.
Inserting this into Eq.~(\ref{respf.5}) and using Eq.~(\ref{nw.2}), we get the third
order response of the first moment
\begin{eqnarray}
& & R^{(1,3)}_{i;j_{1}j_{2}j_{3}}(t,t_{1},t_{2},t_{3})=
\left(\frac{\beta}{2}\right)^{3}\Big\{\frac{\partial^{3}}{\partial
t_{M}\partial t_{I} \partial t_{m}}M_{ij_Mj_Ij_m}(t,t_M,t_I,t_m)  \nonumber \\
& - & \frac{\partial^{2}}{\partial
t_{M}\partial
t_{I}}\langle\hat{\sigma}_{i}(t)\hat{\sigma}_{j_{M}}(t_{M})\hat{\sigma}_{j_{I}}(t_{I})
\hat{B}_{j_{m}}(t_{m})\rangle
- \frac{\partial^{2}}{\partial t_{M}\partial
t_{m}}\langle\hat{\sigma}_{i}(t)\hat{\sigma}_{j_{M}}(t_{M})\hat{B}_{j_{I}}(t_{I})
\hat{\sigma}_{j_{m}}(t_{m})\rangle \nonumber \\
& - & \frac{\partial^{2}}{\partial
t_{I}\partial
t_{m}}\langle\hat{\sigma}_{i}(t)\hat{B}_{j_{M}}(t_{M})\hat{\sigma}_{j_{I}}(t_{I})
\hat{\sigma}_{j_{m}}(t_{m})\rangle
+ \frac{\partial}{\partial
t_{m}}\langle\hat{\sigma}_{i}(t)\hat{B}_{j_{M}}(t_{M})\hat{B}_{j_{I}}(t_{I})
\hat{\sigma}_{j_{m}}(t_{m})\rangle \nonumber \\
& + & \frac{\partial}{\partial
t_{I}}\langle\hat{\sigma}_{i}(t)\hat{B}_{j_{M}}(t_{M})\hat{\sigma}_{j_{I}}(t_{I})
\hat{B}_{j_{m}}(t_{m})\rangle
+ \frac{\partial}{\partial
t_{M}}\langle\hat{\sigma}_{i}(t)\hat{\sigma}_{j_{M}}(t_{M})\hat{B}_{j_{I}}(t_{I})
\hat{B}_{j_{m}}(t_{m})\rangle \nonumber \\
& - &
\langle\sigma_{i}(t)B_{j_{M}}(t_{M})B_{j_{I}}(t_{I})B_{j_{m}}(t_{m})\rangle
\Big\} + \frac{\beta^{3}}{4}\Big\{\Big(\frac{\partial}{\partial
t_{m}}
\langle\hat{\sigma}_{i}(t)\hat{\sigma}_{j_{I}}(t_{I})\hat{B}_{j_I}(t_I)\hat{\sigma}_{j_{m}}(t_{m})\rangle \nonumber \\
& + & \frac{\partial}{\partial t_{m}}\frac{\partial}{\partial
t_{I}} M_{ij_Mj_Ij_m}(t,t_M,t_I,t_m)
-\langle\hat{\sigma}_{i}(t)\hat{\sigma}_{j_{I}}(t_{I})\hat{B}_{j_I}(t_I)\hat{B}_{j_{m}}(t_{m})\rangle
\nonumber \\
& - & \frac{\partial}{\partial t_I}
\langle\hat{\sigma}_{i}(t)\hat{\sigma}_{j_M}(t_M)\hat{\sigma}_{j_{I}}(t_{I})\hat{B}_{j_{m}}(t_{m})\rangle
\Big) \delta_{j_{M}j_{I}}\delta(t_{M}-t_{I}) +
\Big(\frac{\partial}{\partial t_{M}}
\langle\hat{\sigma}_{i}(t)\hat{\sigma}_{j_{M}}(t_{M})\hat{\sigma}_{j_{I}}(t_{I})\hat{B}_{j_I}(t_I)\rangle \nonumber \\
& + & \frac{\partial}{\partial t_{M}}\frac{\partial}{\partial
t_{m}}M_{ij_Mj_Ij_m}(t,t_M,t_I,t_m) -
\langle\hat{\sigma}_{i}(t)\hat{B}_{j_{M}}(t_{M})\hat{\sigma}_{j_{I}}(t_{I})\hat{B}_{j_I}(t_I)\rangle
\nonumber \\
& - & \frac{\partial}{\partial
t_m}\langle\hat{\sigma}_{i}(t)\hat{B}_{j_{M}}(t_{M})\hat{\sigma}_{j_{I}}(t_{I})\hat{\sigma}_{j_m}(t_m)\rangle
\Big) \delta_{j_{I}j_{m}}\delta(t_{I}-t_{m})\Big\} +
\frac{\beta^3}{2}\Big(\frac{\partial}{\partial t_{1}} M_{ij_1}(t,t_1) \nonumber \\
& - & \langle\hat{\sigma}_{i}(t)\hat{B}_{j_{1}}(t_{1})\rangle\Big)
\delta_{j_{1}j_{2}}\delta_{j_{2}j_{3}}\delta(t_{1}-t_{2})\delta(t_{2}-t_{3}).
\label{threekicks}
\end{eqnarray}
At stationarity this becomes
\begin{eqnarray}
R^{(1,3)}_{ij_{1}j_{2}j_{3}}(t,t_{1},t_{2},t_{3})&=&\frac{\beta^{3}}{4}\Big\{
\frac{\partial^{3}}{\partial t_{M}\partial t_{I} \partial
t_{m}}M_{ij_Mj_Ij_m}(t,t_M,t_I,t_m) \nonumber \\
&-&\frac{\partial^{2}}{\partial t_{M}\partial
t_{m}}\langle\hat{\sigma}_{i}(t)\hat{\sigma}_{j_{M}}(t_{M})\hat{B}_{j_{I}}(t_{I})
\hat{\sigma}_{j_{m}}(t_{m})\rangle \nonumber \\
&-&\frac{\partial^{2}}{\partial t_{I}\partial
t_{m}}\langle\hat{\sigma}_{i}(t)\hat{B}_{j_{M}}(t_{M})\hat{\sigma}_{j_{I}}(t_{I})
\hat{\sigma}_{j_{m}}(t_{m})\rangle \nonumber \\
&+& \frac{\partial}{\partial
t_{m}}\langle\hat{\sigma}_{i}(t)\hat{B}_{j_{M}}(t_{M})\hat{B}_{j_{I}}(t_{I})
\hat{\sigma}_{j_{m}}(t_{m})\rangle\Big\} \nonumber \\
&+&\frac{\beta^{3}}{2}\Big\{\frac{\partial}{\partial t_{m}}
\langle\hat{\sigma}_{i}(t)\hat{\sigma}_{j_{I}}(t_{I})\hat{B}_{j_I}(t_I)\hat{\sigma}_{j_{m}}(t_{m})\rangle
\nonumber \\
&+& \frac{\partial}{\partial t_{m}}\frac{\partial}{\partial t_{I}} M_{ij_Mj_Ij_m}(t,t_M,t_I,t_m) 
\Big\}\delta_{j_{M}j_{I}}\delta(t_{M}-t_{I}) \nonumber \\
&+&\beta^3\frac{\partial}{\partial t_{1}} M_{ij_1}(t,t_1)
\delta_{j_{1}j_{2}}\delta_{j_{2}j_{3}}\delta(t_{1}-t_{2})\delta(t_{2}-t_{3}). \nonumber \\
\label{TK9}
\end{eqnarray}
It should be recalled that the singular terms in the last two equations are present only
in the Ising spin case.

\section{Appendix II}

The expansion of the left hand side of  Eq.~(\ref{Ub.16}) can be done in two steps.
Expanding first the exponential we get
\begin{eqnarray}
& & \left \langle \varphi_i(t_F)  \exp \left \{-\beta \int_{t_0}^{t_F} dt \; h(t)\dot{\varphi}(t) \right \}
\right \rangle_{I \rightarrow \beta,[h(t)]} \nonumber \\
& = & \sum_{m=0}^{\infty} {(-\beta)^m \over m!} \sum_{j_1...j_m}
\int_{t_0}^{t_F}dt_1...dt_m \langle \varphi_i(t_F)\dot{\varphi}_{j_1}(t_1)...\dot{\varphi}_{j_m}(t_m)
\rangle_{I \rightarrow \beta,[h(t)]} \nonumber \\
& & h_{j_1}(t_1)...h_{j_m}(t_m)
\label{Ub.18}
\end{eqnarray}
and expanding the individual averages
\begin{eqnarray}
& & \langle \varphi_i(t_F)\dot{\varphi}_{j_1}(t_1)...\dot{\varphi}_{j_m}(t_m)
\rangle_{I \rightarrow \beta,[h(t)]} = \nonumber \\
& & \sum_{p=0}^{\infty} {1 \over p!}  \sum_{q_1...q_p} \int_{t_0}^{t_F}dt_1^{\prime}...dt_p^{\prime}
\left . {\delta^p \langle \varphi_i(t_F)\dot{\varphi}_{j_1}(t_1)...\dot{\varphi}_{j_m}(t_m)
\rangle_{I \rightarrow \beta,[h(t)]} \over \delta h_{q_1}(t_1^{\prime})...\delta h_{q_p}(t_p^{\prime})}
\right |_{h=0} \nonumber \\
& & h_{q_1}(t_1^{\prime})...h_{q_p}(t_p^{\prime})
\label{Ub.19}
\end{eqnarray}
all together these two contributions give
\begin{eqnarray}
& & \left \langle \varphi_i(t_F)  \exp \left \{-\beta \int_{t_0}^{t_F} dt \; h(t)\dot{\varphi}(t) \right \}
\right \rangle_{I \rightarrow \beta,[h(t)]} = \nonumber \\
& & \sum_{m=0}^{\infty}\sum_{p=0}^{\infty}{(-\beta)^m \over m!p!}
\sum_{j_1...j_m} \sum_{q_1...q_p} \int_{t_0}^{t_F}dt_1...dt_m dt_1^{\prime}...dt_p^{\prime}
\left . {\delta^p \langle \varphi_i(t_F)\dot{\varphi}_{j_1}(t_1)...\dot{\varphi}_{j_m}(t_m)
\rangle_{I \rightarrow \beta,[h(t)]} \over \delta h_{q_1}(t_1^{\prime})...\delta h_{q_p}(t_p^{\prime})}
\right |_{h=0} \nonumber \\
& & h_{j_1}(t_1)...h_{j_m}(t_m)h_{q_1}(t_1^{\prime})...h_{q_p}(t_p^{\prime}).
\label{Ub.19bis}
\end{eqnarray}
Reorganizing the double sum 
\be
\sum_{m=0}^{\infty}\sum_{p=0}^{\infty}{(-\beta)^m \over m!p!}\sum_{j_1...j_m} \sum_{q_1...q_p}=
\sum_{n=0}^{\infty}\sum_{p=0}^{n}{(-\beta)^{n-p} \over (n-p)!p!}\sum_{j_1...j_{n-p}} \sum_{q_1...q_p}
\label{Ub.20}
\ee
the above result can be rewritten as
\begin{eqnarray}
& & \left \langle \varphi_i(t_F)  \exp \left \{-\beta \int_{t_0}^{t_F} dt \; h(t)\dot{\varphi}(t) \right \}
\right \rangle_{I \rightarrow \beta,[h(t)]} =  \langle \varphi_i(t_F) \rangle_{I \rightarrow \beta,0} \nonumber \\
& & \sum_{n=1}^{\infty}{1 \over n!} \sum_{p=0}^{n}(-\beta)^{n-p} \left [ {n! \over (n-p)!p!} \right ]
\sum_{j_1...j_{n-p}} \sum_{q_1...q_p}
\int_{t_0}^{t_F}dt_1...dt_{n-p} dt_1^{\prime}...dt_p^{\prime} \nonumber \\
& & \left . {\delta^p \langle \varphi_i(t_F)\dot{\varphi}_{j_1}(t_1)...\dot{\varphi}_{j_{n-p}}(t_{n-p})
\rangle_{I \rightarrow \beta,[h(t)]} \over \delta h_{q_1}(t_1^{\prime})...\delta h_{q_p}(t_p^{\prime})}
\right |_{h=0}
h_{j_1}(t_1)...h_{j_{n-p}}(t_{n-p})h_{q_1}(t_1^{\prime})...h_{q_p}(t_p^{\prime}).\nonumber \\
\label{Ub.21}
\end{eqnarray}
Notice that the combinatorial factor in the square bracket gives the
number of the distinct permutations among the two sets of indeces $(j_1,...,j_{n-p})$ and
$(q_1,...,q_p)$.

Going over to the right hand side of Eq.~(\ref{Ub.16}) and introducing the shorthand
\be
\left \langle P_I(\varphi(t_F)) e^{\beta  {\cal H}_0(\varphi(t_F))}  \varphi_i(t_0)
\right \rangle_{\beta,0 \rightarrow \beta,[\widetilde{h}(t)]} =
\langle RHS \rangle_{\beta,0 \rightarrow \beta,[\widetilde{h}(t)]}
\label{Ub.22}
\ee
one gets
\begin{eqnarray}
& & \langle RHS \rangle_{\beta,0 \rightarrow \beta,[\widetilde{h}(t)]} = \langle RHS \rangle_{\beta,0} \nonumber \\
& + & \sum_{n=1}^{\infty}{1 \over n!} \sum_{j_1...j_n} \int_{t_0}^{t_F}dt_1...dt_{n}
\left . {\delta^n \langle RHS \rangle_{\beta,0 \rightarrow \beta,[\widetilde{h}(t)]}  \over
\delta h_{j_1}(t_1)...\delta h_{j_n}(t_n)} \right |_{h=0} \nonumber \\
& & h_{j_1}(t_1)... h_{j_n}(t_n)
\label{Ub.23}
\end{eqnarray}
and, since $\widetilde{h}_j(\widetilde{t}) = h_j(t)$,
this can be rewritten as
\begin{eqnarray}
& & \langle RHS \rangle_{\beta,0 \rightarrow \beta,[\widetilde{h}(t)]} = \langle RHS \rangle_{\beta,0} \nonumber \\
& + & \sum_{n=1}^{\infty}{1 \over n!} \sum_{j_1...j_n} \int_{t_0}^{t_F}dt_1...dt_{n}
\left . {\delta^n \langle RHS \rangle_{\beta,0 \rightarrow \beta,[\widetilde{h}(t)]}  \over
\delta \widetilde{h}_{j_1}(\widetilde{t}_1)...\delta\widetilde{h}_{j_n}(\widetilde{t}_n)} \right |_{h=0} \nonumber \\
& & h_{j_1}(t_1)... h_{j_n}(t_n)
\label{Ub.25}
\end{eqnarray}
where $\langle \cdot \rangle_{\beta,0}$ stands for the equilibrium average at the temperature $\beta$
and without external field.
Therefore, comparing with Eq.~(\ref{Ub.21}), one arrives at Eqs.~(\ref{Ub.26}) and~(\ref{Ub.27}).


\begin{thebibliography}{99}
\bibitem{nota1} See for instance Ref.~\cite{semerjian}, where the derivation of the Onsager
relation is carried out for continous variables, but it works exactly in the same
way also for Ising spins.
\bibitem{noi}
The actual value of this exponent is controversial. Our results can be found in Ref.~\cite{expa,expab}.
What is uncontroversial, and what matters here, is that for $d>1$ and for Ising spins $a>0$.
\end{thebibliography}
\end{document}